\documentclass[english]{article}

\usepackage{geometry}
\usepackage{rotating}
\usepackage{color}
\usepackage{graphicx}
\usepackage{graphics}
\usepackage{epstopdf}
\usepackage{setspace}
\usepackage[authoryear]{natbib}
\usepackage{xfrac}
\usepackage{algpseudocode}
\usepackage{algorithm}
\usepackage{amsmath}
\usepackage{comment}
\usepackage{url}
\usepackage[absolute]{textpos}
\usepackage{babel}
\usepackage{multicol,caption}
\usepackage{wrapfig}


\begin{document}

\begin{textblock}{5}(11,1) 
\end{textblock}

\title{\Large \bf Robust forward simulations of recurrent hitchhiking}

\author{Lawrence H. Uricchio$^{1}$, Ryan D. Hernandez$^{2,3,4}$}

\date{\today}

\maketitle
\noindent $^{1}$UC Berkeley \& UCSF Joint Graduate Group in Bioengineering, San Francisco, CA 
\noindent $^{2}$Department of Bioengineering and Therapeutic Sciences,
\noindent $^{3}$Institute for Human Genetics,
\noindent $^{4}$Institute for Quantitative Biosciences (QB3), UCSF, San Francisco, CA
~\\
\\
\\
\\
\\
\\
\\
\\
\\
\\
\\
\\
\\
\\
\\
\\
\clearpage
~\\
{\Large \bf Simulations of recurrent selection}
\\
\\
\noindent Corresponding author: Ryan D. Hernandez, ryan.hernandez@ucsf.edu
\\
\\
UCSF
\\
Box 2530
\\
1700 4th Street
\\
San Francisco, CA 94158
\\
Phone: 415-514-9816
\\ 
Fax: 415-514-1028
\\
\clearpage
\begin{abstract}
{\normalsize 
Evolutionary forces shape patterns of genetic diversity within 
populations and contribute to phenotypic variation.  In particular, recurrent positive selection
has attracted significant interest in both theoretical and empirical studies. 
However, most existing theoretical models of recurrent positive selection
cannot easily incorporate realistic 
confounding effects such as interference between selected sites, arbitrary selection schemes, and 
complicated demographic processes.  It is possible to quantify the effects of arbitrarily complex 
evolutionary models by performing forward population genetic simulations, but forward simulations 
can be 
computationally prohibitive for large population sizes ($>10^5$). A common approach for overcoming 
these computational limitations is rescaling of the most computationally expensive parameters, 
especially population size.  Here, we show that \textit{ad hoc} 
approaches to parameter rescaling under the recurrent hitchhiking model do not always provide 
sufficiently accurate dynamics, potentially skewing patterns of diversity in simulated DNA 
sequences.  We derive an extension of the recurrent hitchhiking model that is appropriate for 
strong selection in small population sizes, and use it to develop a method for parameter rescaling 
that provides the best possible computational performance for a given error tolerance.  We perform 
a detailed theoretical analysis of the robustness of rescaling across the parameter space.  
Finally, we apply our rescaling algorithms to
parameters that were previously inferred for \textit{Drosophila}, and discuss practical 
considerations such as interference between selected sites.
}{\normalsize \par}
\end{abstract}

\clearpage

\section{Introduction}

A central goal of population genetics is to determine the strength and rate of natural selection
in populations. Natural selection impacts patterns of genetic diversity within populations,
and is likely to influence phenotypes of biological and medical interest 
(\citealt{Bustamante:2005:Nature:16237444,Torgerson:2009:PLoS-Genet:19662163,
Maher:2012:Hum-Hered:23594490,Arbiza:2013:Nat-Genet:23749186}). There exists a large body 
of literature focused on mathematical models of selection in populations and inferring the action 
of selection on DNA sequences under these models (recent reviews include 
\citealt{Pool:2010:Genome-Res:20067940,Crisci01032012,Cutter:2013:Nat-Rev-Genet:23478346}). 
One such model is known as recurrent hitchhiking, in which patterns of diversity at a selectively 
neutral locus are altered due to repeated positive selection at linked loci.

Recurrent hitchhiking has been theoretically explored (\citealt{GRH:1754360,
Ota:1975:Genet-Res:1183812,Kaplan:1989:Genetics:2612899,Stephan:2006:Genetics:16452153,
Coop:2012:Genetics:22714413}) and applied to DNA sequences 
of various organisms (\citealt{Bachtrog:2008:BMC-Evol-Biol:19091130,
Jensen:2008:PLoS-Genet:18802463,Ingvarsson:2010:Mol-Biol-Evol:19837657,
Singh:2013:Genetics:23105013}).
The classic work of \citet{Stephan1992237} modeled the dynamics of the 
neutral locus in a single sweep with diffusion-based differential equations, which they solved 
approximately. \citet{Wiehe:1993:Mol-Biol-Evol:8355603} later showed that their solution for 
single sweeps could be applied to a recurrent sweep model, where the expected reduction in neutral 
diversity is well approximated by $\frac{r}{r + \alpha I \lambda}$; $\alpha=2Ns$ where $N$ is the 
population size and $s$ is the selection coefficient, $r$ is the recombination rate, $\lambda$ is 
the rate of positively selected substitutions, and $I$ is a constant that approximates the value 
of an integral. However, 
little work has been done to explore recurrent sweeps with forward simulations (but see 
\citealt{Kim:2003:Genetics:12750349}, and \citealt{Chevin2008}, where interfering substitutions 
were studied with forward simulations, and the discussion herein).

It is crucial to understand the dynamics of recurrent sweeps (and other population genetic models) 
when realistic perturbations to the model are introduced, which is often difficult in a coalescent
framework. In contrast, with forward simulations it is straightforward to introduce arbitrarily 
complex models, including demographic processes, interference between selected 
sites, simultaneous negative and positive selection, and variable strength of selection or 
recombination rate across a chromosome. Furthermore, forward simulations can be performed exactly 
under a given model, and hence they can be used as a direct test of theoretical 
predictions.  Simulations can be used in conjunction with inference methods such as
Approximate Bayesian Computation to estimate parameters when the likelihood function of the
data under the model is unknown
(\citealt{Beaumont:2002:Genetics:12524368}).

In population genetics, forward methods have often been overlooked in favor of reverse time 
coalescent simulators due to computational efficiency 
(\citealt{Hernandez:2008:Bioinformatics:18842601}; for an overview of coalescent and forward 
simulation techniques, see \citealt{Kim01012009}).  Although coalescent simulations are generally 
more computationally efficient, in most applications they require some \textit{a priori} knowledge 
of allele trajectories. Recent improvements in computer memory and processor speeds have made 
forward simulations more tractable. However, simulations of recurrent hitchhiking in some parameter 
regimes of interest (e.g., $N>10^5$) are still computationally prohibitive, so it is frequently 
necessary to rescale model parameters (e.g., $N$ and chromosome length, $L$) 
(\citealt{Kim01012009}). Currently, the literature provides some guidelines for 
performing parameter scaling in forward simulations (\citealt{Hoggart:2007:Genetics:17947444}), but
it is not clear that these methods will be generally applicable to all models or hold in
all parameter regimes.

In this investigation, we examine recurrent sweeps through forward simulation and theory.  
We provide a detailed, practical discussion of simulations of
recurrent sweeps in a forward context, focusing on scaling laws of relevant parameters such as 
$N$, $\lambda$, $r$, $\alpha$, and $L$.  We evaluate a ``naive'' parameter rescaling algorithm, 
and show that this technique can bias patterns
of variation in the simulations because it is not conservative with respect to the underlying
genealogical process, particularly in the large $\alpha$, small $N$ regime.  
We quantify the effect of large values of the 
selection coefficient $s$ on recurrent hitchhiking through theory. Finally, we leverage these 
principles to make gains in computational efficiency with a simple algorithm that provides the best
possible performance for a prespecified error threshold, and apply the method to simulations of 
parameters previously inferred in \textit{Drosophila}.

\section{Model}

Here, we describe the recurrent hitchhiking model (shown schematically in Figure 1), 
upon which we build the results and simulations 
in this article.  Key parameters of the model are discussed below, and summarized in Table 1. 

A neutral locus is flanked on both sides by sequences experiencing repeated positively selected 
substitutions at rate $\lambda$ per generation per site. 
$\lambda$ is assumed to be small enough that multiple positively selected mutations do
not simultaneously sweep in the population and hence there is no interference (though interference
between selected sites is not prohibited in the simulations performed herein).
Population size is fixed at $N$.  In 
forward simulations, there is no distinction between effective and census population size, so 
$N=N_e$. Recombination occurs at rate $r_f$ (recombination fraction) per generation per 
chromosome. Note that the recombination fraction is the probability that the number of 
recombination events between two loci is an odd number, and cannot exceed 0.5.
Each positively selected site has a selection coefficient 
$s=\sfrac{\alpha}{2N}$.  Heterozygous individuals have fitness $1+s$, while individuals homozygous 
for the selected allele have fitness $1+2s$. The neutral locus itself is assumed to be 
non-recombining. Any of the above constraints and model assumptions can be relaxed in forward 
simulations.

\linespread{1.2}
\begin{table*}[ht]
    \caption{\textbf{Parameter definitions}} 
    \centering  
    \begin{tabular}{c l} 
     \hline\hline

     $N$ & Population size \\
     $\tau_f$ & Time of fixation of selected allele \\
     $t$ & Time in generations \\
     $x(t)$ & Frequency of the selected allele \\
     $h(t)$ & Relative heterozygosity at the neutral locus among selected chromosomes \\
     $q_l$ & Minor allele frequency at site $l$ \\
     $l_0$ & Length of neutral region \\
     $L$ & Flanking sequence length \\
     $\pi$ & Nucleotide diversity, $\pi = \sum_{l=1}^{l_0} \frac{2q_l(1-q_l)(n-1)}{n}$ \\
     $\pi_0$ & Nucleotide diversity under neutrality \\
     $\pi_N$ & Nucleotide diversity in a population of size $N$ \\
     $p(t)$ & Probability of common ancestry at the neutral locus \\
     $n$ & Number of sampled sequences \\
     $\mu$ & Mutation rate/generation/chromosome/base pair \\
     $\theta=4N\mu$ & Population scaled mutation rate \\
     $s$ & Selection coefficient \\ 
     $\alpha=2Ns$ & Population scaled selection strength\\
     $r$ & Recombination rate/generation/chromosome/base pair \\
     $\rho=4Nr$ & Population scaled recombination rate \\
     $r_f$ & Recombination fraction (probability of productive recombination, 
       per generation per chromosome) \\
     $\lambda$ & Rate of positively selected substitutions per site per generation \\
     $k_h$ & Rate of common ancestry induced by sweep events (see \eqref{eq:kh1})\\  
     $R_c$ & Total rate of common ancestry induced by sweep and coalescent events \\  
     $I_{\alpha,s}$ & The integral $\int_0^{u*}p_{\tau_f}(u)du$ (see \eqref{eq:kh1})\\ 
     $I^{*}_{\alpha,s}$ & The integral $\int_0^{u*}p^{*}_{\tau_f}(u)du$ (see \eqref{eq:Epistar})\\ 
     $I=0.075$ & A constant approximating $I_{\alpha,s}$\\ [1ex]
    \hline 
\end{tabular}
\label{table:params} 
\end{table*}
\linespread{2}

Consider the coalescent history at the neutral locus of two sequences sampled immediately after a 
selective sweep. If there is no recombination between the neutral and selected loci during the 
sweep, then the two sequences must share a common ancestor at some point during the sweep. If 
selection is sufficiently strong, then the time to fixation of selected alleles is effectively 
instantaneous relative to the neutral fixation process (\citealt{Kaplan:1989:Genetics:2612899}), 
and thus the expected heterozygosity at the neutral site at the completion of the sweep is nearly 
0 because very few mutations are introduced during the sweep.

Recombination significantly complicates this model.  Immediately after a sweep, the reduction in 
heterozygosity at the neutral locus is a function of the recombination distance between the 
selected substitution and the neutral locus, and the strength of selection.  Stephan, Wiehe, and 
Lenz (SWL) calculated the reduction in heterozygosity at the neutral locus with a diffusion based, 
differential equation framework (\citealt{Stephan1992237}). They showed that the expected reduction 
in heterozygosity at the neutral locus among chromosomes carrying the selected allele, relative to 
the baseline heterozygosity, $h(t)$, can be modeled with a simple differential equation, which they
solved approximately.

\citet{Kaplan:1989:Genetics:2612899} showed that 
$h(t)$ is closely related to the probability that two sequences sampled at the end of the sweep
share a common ancestor at the neutral site during the sweep,\ $p(t)$.
\begin{equation}
\label{eq:ph}
p(t) = 1-h(t)
\end{equation}
\noindent
This allows the results of SWL to be interpreted in terms of the coalescent process at the neutral 
locus.   
Note that when 
$t=\tau_f$ (the end of the sweep), $p(\tau_f)$ represents the probability of common ancestry at the
neutral locus for a pair of sequences at some point during the sweep
because all chromosomes carry the selected allele at the end of a sweep. Throughout the article, we 
subscript variables of interest with $\tau_f$ to denote their values at the time of fixation and 
emphasize their dependence on the recombination fraction $r_f$ (e.g., $p_{\tau_f}(r_f)$). 
Rewriting SWL results with \eqref{eq:ph}, we obtain

\begin{equation}
\label{eq:pt}
\frac{d}{dt}p(t) = \frac{1-p(t)}{2Nx(t)}  - 2r_f\ p(t)\ (1-x(t))
\end{equation}
\noindent
where $x(t)$ is the frequency of the selected allele at time $t$ during the sweep.
Equation \eqref{eq:pt} is equivalent to equation 5 of \citet{GRH:3537} when the selected allele is 
at low frequency.  

Equation \eqref{eq:pt} can be interpreted in terms of the recombination process between the 
neutral and selected loci. In particular, there are two mechanisms that can change the proportion 
of selected sequences that share common ancestry at the neutral locus. The first term on the RHS of
\eqref{eq:pt} represents that chance of common ancestry in the previous generation among selected 
sequences that have already recombined off of the original background. The chance that any two such 
sequences share a common ancestor in the previous generation is $\frac{1}{2Nx(t)}$. The second 
term represents the chance that a recombination event occurs between a selected chromosome and some
non-selected chromosome, thereby reducing $p(t)$.  The first term is only important when the 
frequency of the selected site is low, because it is inversely proportional to the number of 
selected chromosomes, whereas the second term contributes non-negligibly to the dynamics at all 
allele frequencies of the selected locus.

Consider the coalescent history at the neutral locus of two lineages sampled at the current time 
(not necessarily immediately after a sweep event).  In each
preceding generation, there is some chance that they share a common ancestor at the neutral locus
due to normal coalescent events, and some chance that they share common ancestry because of a sweep 
event.  Since sweeps occur nearly instantaneously relative to the timescale of coalescence under 
neutrality, we can approximate the chance of common ancestry as two competing processes. Neutral 
events occur at rate $\sfrac{1}{2N}$ and compete with sweep events, which happen at rate 
$\frac{2\lambda}{r}p_{\tau_f}(r_f)dr_f$ in a window of size $dr_f$, assuming that sweeps occur 
homogeneously across the chromosome and $r_f \approx rL$.  Note that $r$ and $\lambda$ appear in
a quotient in this rate, which implies that multiplying both the substitution and 
recombination rates by a common factor has no impact on the model. 
The factor of 2 represents the flanking sequence on either side of the neutral locus.

Following the results of SWL, an approximate solution to \eqref{eq:pt} is:
\begin{equation}
\label{eq:ptau}
p_{\tau_f}(r_f) = 
1- \frac{2r_f}{s} \alpha^{\frac{-2r_f}{s}} \Gamma\left[\frac{-2r_f}{s},\frac{1}{\alpha}\right]
\end{equation}
\noindent
where $\Gamma$ is the incomplete gamma function. Note that \eqref{eq:ptau} is a function of $r_f, 
\alpha$, and $s$, but we only denote the dependence on $r_f$ since $\alpha$ and $s$ are assumed to 
be fixed for the analysis herein.

Following SWL, we denote the rate at which lineages merge due to sweep events as $k_h$,

\begin{equation}
\label{eq:kh1}
k_h = 2N\left(\frac{2\lambda}{r}\int_0^{{r_f}^{*}}p_{\tau_f}(r_f)\,dr_f\right)
\end{equation}
\noindent
where ${r_f}^{*}$ is taken as the value of $r_f$ that corresponds to the end of the flanking 
sequence.  If the flanking chromosome being modeled exceeds $\frac{s}{r}$ base pairs, 
previous work suggests that $r_f^{*}$ can be taken to be any value sufficiently far away from the 
neutral locus such that the value of $k_h$ is as close as desired to its asymptotic limit 
\citep{Jensen:2008:PLoS-Genet:18802463}. The factor of $2N$ is introduced 
to rescale in coalescent units, such that neutral coalescent events happen at rate $1$ relative to 
sweep merger events. 

\subsection{The expectation of $\pi$ in recurrent hitchhiking}

In the recurrent hitchhiking (RHH) model, it is of great interest to describe the reduction in 
diversity as a function of the basic parameters of the model ($\alpha, r, \lambda$, etc.).  To make 
this dependence clearer, we perform two changes of variables in \eqref{eq:kh1}. First, we note that 
\eqref{eq:kh1} was derived by SWL under the assumption that $r_f^{*}$ is small, such that the 
recombination fraction is given by $r_f \approx rL$.  Here we will frequently be concerned with 
values of $r_f$ that approach its maximum value of 0.5, which invalidates this approximation.  We 
therefore rewrite \eqref{eq:kh1} as a function of $L$, substituting 
$r_f = \frac{1-e^{-2rL}}{2}$ for the quantity $r_f$ (\citealt{haldane1919}).  We then substitute 
the quantity $u = \frac{2r}{s}L$ for $L$. Rewriting $p_{\tau_f}$ and $k_h$ as functions of $u$, we 
have 

\begin{equation}
\label{eq:ptauu}
p_{\tau_f}(u) = 1 - \frac{(1-e^{-su})}{s}\alpha^{\frac{-(1-e^{-su})}{s}}\Gamma\left[\frac{-(1-
e^{-su})}{s},\frac{1}{\alpha}\right]
\end{equation}  

and 

\begin{equation}
\label{eq:kh2}
k_h = \frac{2Ns\lambda}{r}\int_0^{{u}^{*}}p_{\tau_f}(u)\,du
\end{equation}

It is useful to examine the properties of \eqref{eq:ptauu} and \eqref{eq:kh2} as a 
function of $s$. When $s$ is small, \eqref{eq:ptauu} can be rewritten as 

\begin{equation}
p_{\tau_f}(u) \approx 1 - u\alpha^{-u}\ \Gamma\left[-u,\frac{1}{\alpha}\right]
\end{equation} 
\noindent
which removes the dependence on $s$ and is identical to the quantity inside the integral on the RHS
of equation 4 of \citet{Wiehe:1993:Mol-Biol-Evol:8355603}.  Thus, the integral on the RHS 
of \eqref{eq:kh2} is a function only of the parameter $\alpha$ when $s$ is small. 
This is not necessarily the case as $s$ becomes large, but we also note that \eqref{eq:ptau} was 
originally derived under the assumption that $s$ is small, so it is possible that the large $s$ 
behavior is not accurately captured by \eqref{eq:ptauu} and \eqref{eq:kh2}. 

Similar to \citet{Wiehe:1993:Mol-Biol-Evol:8355603}, we define the integral in \eqref{eq:kh2} 
as $I_{\alpha,s}$, but we include the subscript $\alpha,s$ to emphasize that, under some 
circumstances, $I_{\alpha,s}$ may be a function of both $\alpha$ and $s$ and cannot be written
as a function of only the population scaled strength of selection. The total rate of 
coalescence $R_c$ (in coalescent units) due to both sweep and neutral coalescent events is then
\begin{equation}
\label{eq:Rc}
R_{c} = 1 + k_h = 1 + \frac{\alpha\lambda}{r}I_{\alpha,s}
\end{equation}
\noindent
The expected height of the coalescent tree for two sequences is the inverse of this rate.  The 
expected reduction in diversity at the neutral locus is proportional to the decrease in the
height of the coalescent tree, relative to neutrality.  
\begin{equation}
\label{eq:Epi}
E_{\alpha,s}\left[\sfrac{\pi}{\pi_0}\right]= \frac{1}{R_c} = 
\frac{r}{r+\alpha\lambda I_{\alpha,s}}
\end{equation}
\noindent
\citet{Wiehe:1993:Mol-Biol-Evol:8355603} found that $I_{\alpha,s}$ is approximately constant 
($I$=0.075) over a range of large values of $\alpha$.  

\begin{equation}
\label{eq:EpiWS}
E_{WS}\left[\sfrac{\pi}{\pi_0}\right]= 
\frac{r}{r+\alpha\lambda I}
\end{equation}
\noindent
Note that this removes the dependence on $s$, which is asserted 
by \eqref{eq:Epi}. In the following sections we show that both \eqref{eq:Epi} and \eqref{eq:EpiWS}
may not hold when $s$ is large.

\section{Materials \& Methods}

\subsection{Simulating RHH Models}

We performed forward simulations of RHH with SFS\_CODE (\citealt{Hernandez:2008:Bioinformatics:18842601}; 
see Appendix for details). A pictorial representation of the model is shown in Figure
1. 

\begin{figure*}[t]
  \includegraphics[height=240pt,keepaspectratio=true]{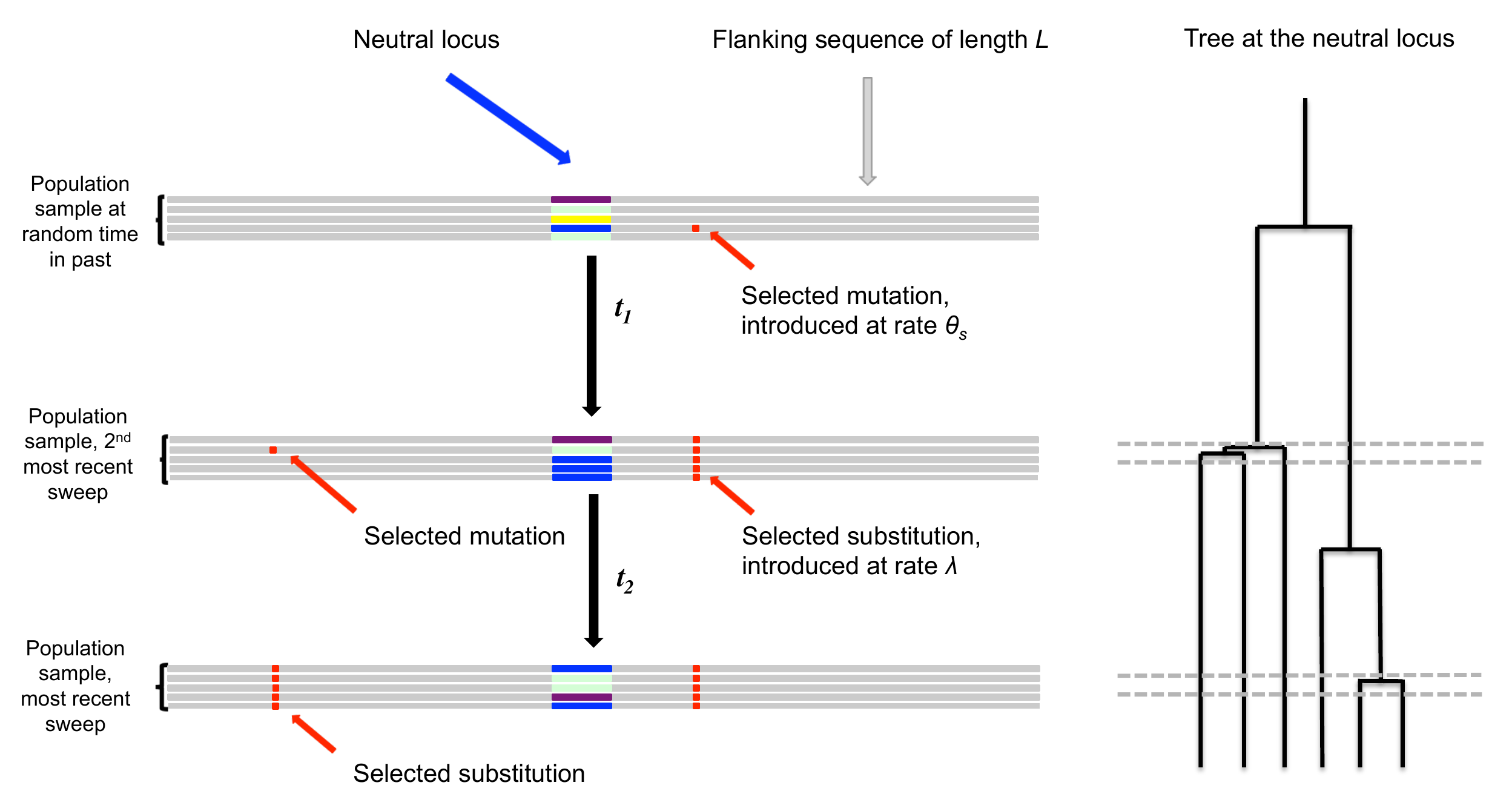}
    \caption{A pictorial representation of the recurrent hitchhiking model. Diverse neutral haplotypes 
     are indicated with various colors at the neutral locus.  When a selected
     site is introduced and eventually goes to fixation, it drags linked neutral variation to higher 
     frequency.  Selected mutations can occur at any distance from the neutral locus within the flanking
     sequence.  
     Viewed from the perspective of the 
     genealogy, sweep events generate an excess of recent common ancestry at a linked neutral site, reducing the overall
     height of the coalescent tree relative to neutrality. Selected sites that are more closely linked to the neutral site have 
     a stronger impact on the overall height of the tree because they induce more common ancestry on a short time scale.
     The overall impact of linked 
     selection at a neutral locus is a function of the strength of selection, the rate of recombination,
     and the rate at which selected sites reach fixation. The neutral site is assumed to be non-recombining, but this assumption
     can be relaxed in simulations.}
\end{figure*}

All simulations in this article were performed with $\theta=0.002$ at the neutral locus, and
reductions in diversity were calculated as the ratio of the observed diversity to $0.002$, unless 
otherwise noted.  Nucleotide diversity $\pi$ and Tajima's $D$ were calculated with a custom script.  
We often report the difference proportion in $\pi_{N_1}$ (diversity in a rescaled population of size $N_1$)
as compared to $\pi_{N_0}$ (diversity in the population of size $N_0$), which we define as 
$\frac{\pi_{N_1}-\pi_{N_0}}{\pi_{N_0}}$.  For each simulation we sampled 10 individuals (20 chromosomes) from 
the population. The neutral loci in all simulations are 1 Kb in length.

\subsection{Fixing the probability of fixation}

Throughout the article, we discuss appropriate choices of $r$, $\lambda$, $s$, $L$, and $N$ for
simulations. However, in forward simulations, the rate of substitution is not explicitly provided
to the software, but rather a rate of mutation. In order to calculate the appropriate
mutation rate for a simulation, one must incorporate the probability of fixation for a positively
selected site.  For $s < 0.1$, the fixation probability of \citet{KIMURA:1962:Genetics:14456043} is 
sufficient:

\begin{equation}
P_{Kimura}(s,\alpha) = \frac{1-e^{-2s}}{1-e^{-2\alpha}}
\end{equation}
\noindent
However, when $s > 0.1$, this approximation overestimates the probability of fixation.  
For $s > 0.1$, we
treat the initial trajectory of the selected site as a Galton-Watson process and calculate the
probability of extinction by generation $i$, $P_e(i)$, with procedure $P_{GW}(s)$ \citep{fisher1999genetical}.
\\
\begin{algorithm}
\begin{algorithmic}
\Procedure{$P_{GW}$}{$s$}
\State $P_e(0) = e^{-(1+s)}$
\State $i = 1$
\While{$P_e(i)-P_e(i-1) > \delta$}
    \State $P_e(i) = e^{-(1+s)(1-P_e(i-1))}$
    \State $i \leftarrow i+1$
\EndWhile \\
\State \textbf{return} $1-P_e(i)$
\EndProcedure
\end{algorithmic}
\end{algorithm}
In practice, this algorithm takes fewer than 200 iterations to converge for $s > 0.1$, and provides
accurate results (Figure 2).  Simulations for this figure were performed with a simple 
Wright-Fisher simulator that only sampled the trajectory of the selected site and followed it until 
either 1) loss or 2) the frequency of the selected site exceeded $\sfrac{100}{\alpha}$, which very 
nearly guarantees eventual fixation.

All analysis and simulation scripts used in this article are available upon request from the
authors.

\begin{figure*}[t]
  \includegraphics[height=285pt,keepaspectratio=true]{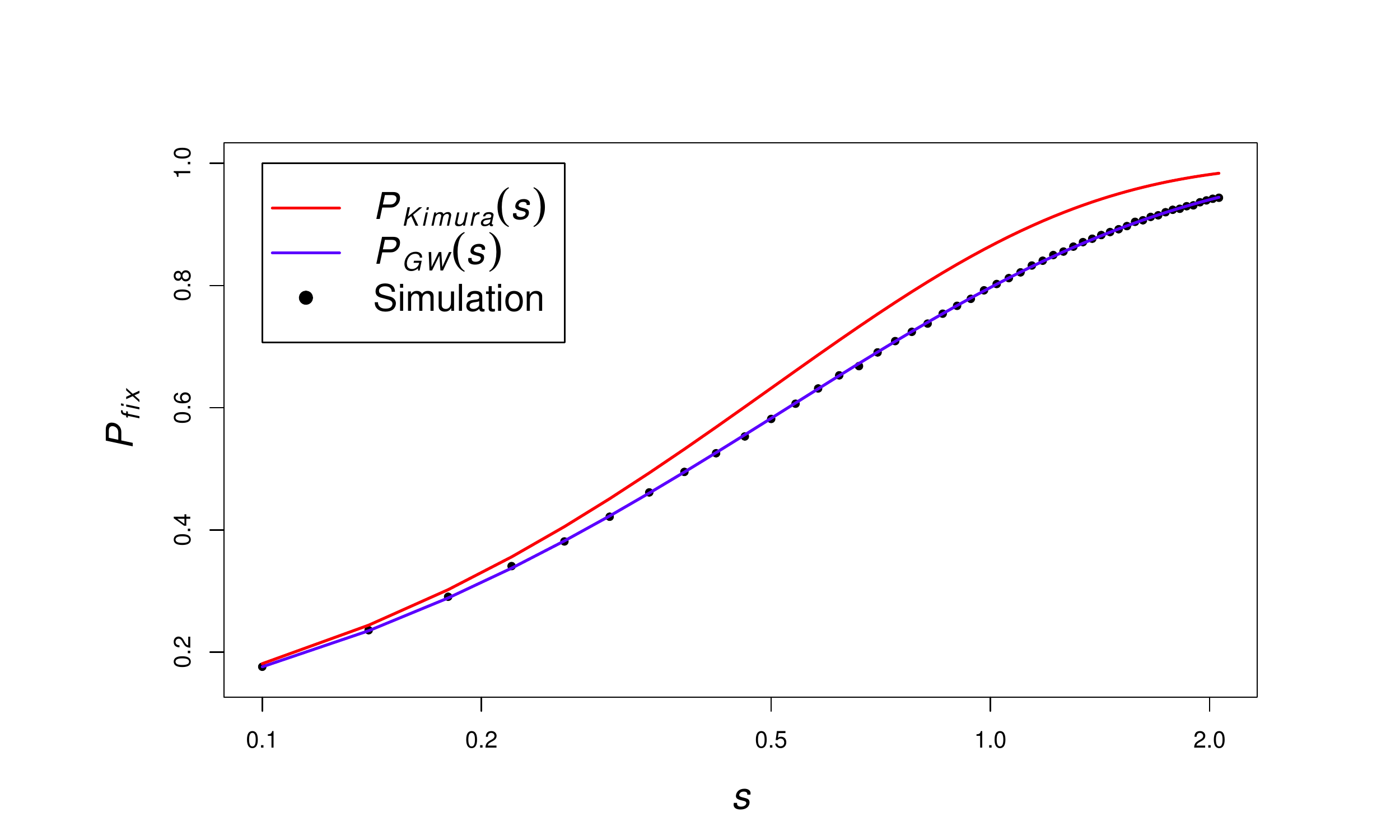}
    \caption{The probability of fixation ($P_{fix}$) as a function of $s$. 
             Simulation points represent the fraction
             of fixations in $10^5$ simulations.  $N=10^4$.}
\end{figure*}

\section{Results}

\subsection{A ``naive'' approach to parameter rescaling}

In forward simulations, the most computationally costly parameters are $N$ and $L$, so we seek to 
reduce these parameters as much as possible. A simple and widely used rescaling assumption is that 
patterns of diversity are conserved when population scaled parameters $\rho=4Nr$, $\alpha=2Ns$, and 
$\theta=4N\mu$ are held fixed and $N$ is varied (\citealt{Kim01012009}; for more discussion, 
see section 5 of the SFS\_CODE manual).  This is equivalent to the statement that the effective 
population size is not a fundamental parameter of the dynamics, and is similar to the rescaling 
strategy described in \citet{Hoggart:2007:Genetics:17947444}, which was not designed specifically 
for RHH simulations.

Equation \eqref{eq:Epi} provides an informed view of rescaling that incorporates RHH theory. 
Equation \eqref{eq:Epi} predicts 
that the impact of the underlying genealogical process on neutral sequence depends on the compound 
parameters $Ns$ and $\sfrac{r}{\lambda}$, but not directly on $\rho$.  Hence, if $s$ is increased 
and $N$ decreased while holding their product constant, and $r$ and $\lambda$ are increased while 
holding their ratio constant, \eqref{eq:Epi} predicts that patterns of variation will be 
maintained.

Finally, \eqref{eq:ptau} suggests as we decrease $N$ and increase $s$ for fixed $\alpha$ we must 
also increase the length of the flanking sequence, because selection at more distant sites can 
impact the neutral locus as $s$ is increased. Note that we can accomplish 
this either by fixing the recombination rate and increasing the length in base pairs of the 
flanking region or by increasing the recombination rate for some fixed flanking length.  
Since the same number of mutations are introduced, these options are functionally identical, 
but the latter may require less RAM for some forward simulation implementations.

Taken together, these scaling principles suggest a simple algorithm for choosing simulated 
values of $N_1$, $\alpha_1$, $r_1$, and $\lambda_1$ that are conservative with respect to 
the genealogical process as predicted by \eqref{eq:Epi}.  We wish to model 
$L_0$ flanking base pairs of sequence in a population of size $N_0$ with parameters $\rho_0, 
\alpha_0$, and 
$\lambda_0$. We choose $L_1$ and $N_1$ to be any computationally convenient flanking length and 
population size.  We compute the remaining simulation parameters with Algorithm 1.

\begin{algorithm}
\begin{algorithmic}
\Procedure{Algorithm 1}{$\rho_0,\alpha_0,L_0,\lambda_0,N_0,N_1,L_1$}
\State Let $s_0 = \frac{\alpha_0}{2N_0}$; $r_0 = \frac{\rho_0}{4N_0}$; $a = \frac{s_0}{L_0r_0}$
\State $\alpha_1 = \alpha_0$
\State $s_1 = \frac{\alpha_1}{2N_1}$
\State $r_1 = \frac{s_1}{aL_1}$
\State $\lambda_1 = \frac{r_1\lambda_0}{r_0}$ 
\State $\rho_1 = 4N_1r_1$
\State \textbf{return} $N_1,\rho_1,\alpha_1,L_1,\lambda_1$
\EndProcedure
\end{algorithmic}
\end{algorithm}

Note that if we choose $L_1=L_0$, we obtain $\alpha_1 = \alpha_0$, $\rho_1 = \rho_0$, and $4N_1\lambda_1 =
4N_0\lambda_0$, which is consistent with the rescaling strategy of \citet{Hoggart:2007:Genetics:17947444}
and diffusion theory.

In Figures 3A-C, we show results obtained with Algorithm 1. In 3A, we plot the normalized 
difference in mean diversity between simulations performed in a population with $N_0=5,000$ and 
simulations performed with rescaled parameters and varying choices of $N_1$.  The
dashed black line at 0 represents the expectation under perfect rescaling, because perfect 
rescaling will result in a normalized difference in means equal to zero between rescaled 
parameters and the 
original parameters. The colored points each represent the mean of 5,000 simulations and the solid 
colored curves were explicitly calculated with \eqref{eq:Epi}. 

Algorithm 1 generates patterns of diversity in the rescaled populations (colored 
points, 3A) that are similar to the simulated diversity in the model population (black dashed line, 
3A) when the strength of selection is low, but the algorithm performs poorly when the strength of 
selection gets arbitrarily large. Qualitatively similar results are observed for the variance in $\pi$ (3B) and 
Tajima's $D$ (3C).  Furthermore, the mean diversity of simulations performed with 
Algorithm 1 does not agree well with explicit calculation of the expected diversity using 
\eqref{eq:Epi} when selection is strong, as seen by the divergence between the mean diversity in 
the simulations and the solid curves.  In fact, \eqref{eq:Epi} predicts that the diversity will 
decrease as $N$ grows because of the dependence of $I_{\alpha,s}$ on $s$ (3A, solid colored 
curves), but simulations show the opposite pattern.  This demonstrates that the simulated value of $s$ 
has some effect on the expected patterns of diversity (which is not predicted by 
the results of \citet{Wiehe:1993:Mol-Biol-Evol:8355603}, which we used to build Algorithm 1), and 
that \eqref{eq:Epi} does not appropriately model this dependence.  

In the next sections we examine circumstances under which the assumptions used to 
derive \eqref{eq:Epi} and \eqref{eq:EpiWS} may break down, and we use insights from this analysis 
to design a more robust approach to parameter rescaling.

\begin{figure*}[t]
  \includegraphics[height=290pt,keepaspectratio=true]{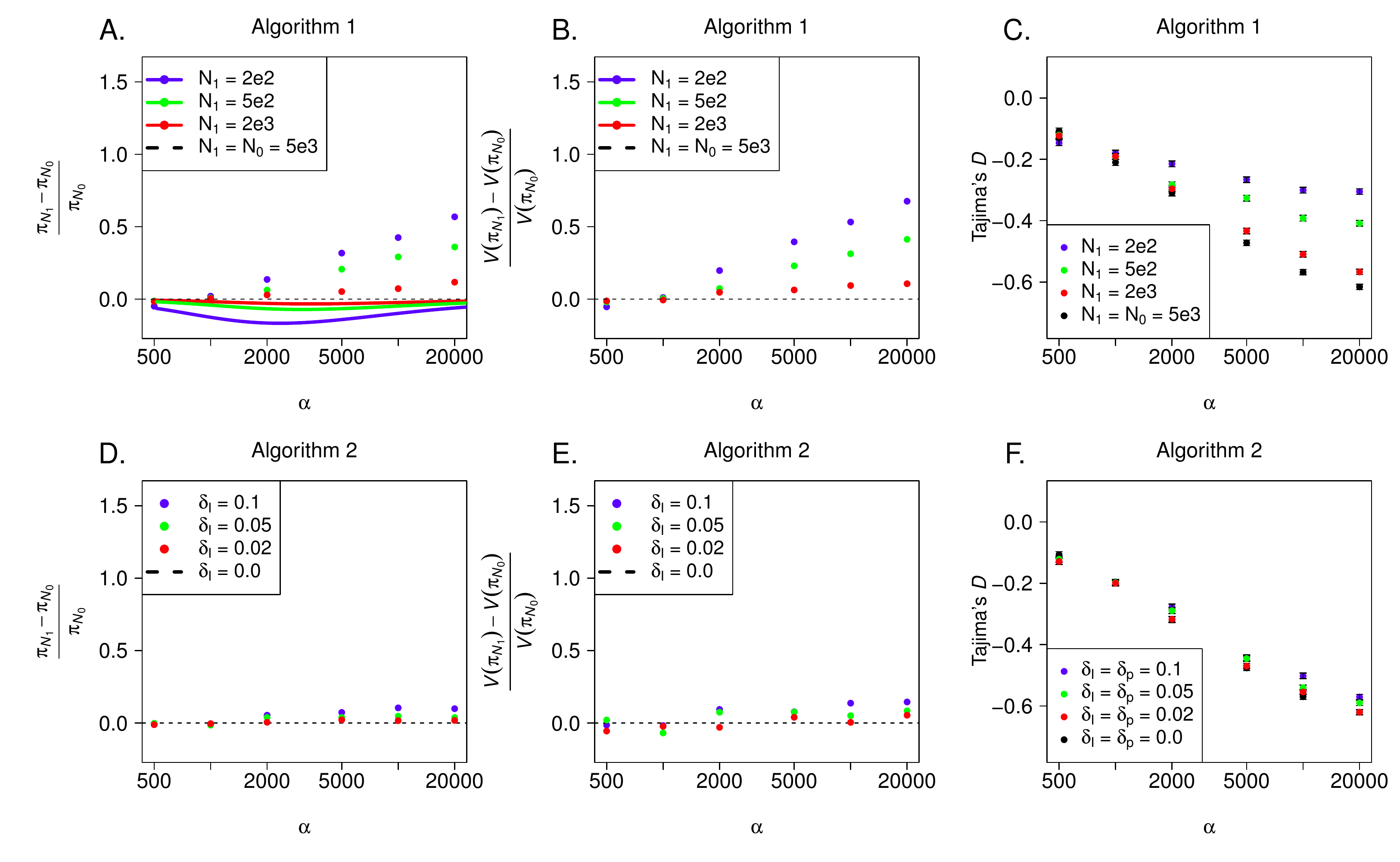}
    \caption{Mean and variance of observed diversity in the rescaled populations ($N_1$) relative to the model 
     population ($N_0=5,000$).
     Rescaled parameters were obtained with Algorithm 1 in panels A, B, and C and with Algorithm 2 in panels D, E, and F. 
     10,000 simulations were performed for each parameter combination.
     The theoretical curves in A were calculated with \eqref{eq:Epi} in \textit{Mathematica}
     (\citealt{Mathemat}).
     Parameters: $N_0 = 5 \times 10^3$, $\rho_0=10^{-3}$, $\lambda_0=10^{-10}$, 
     $L_0=10^6$, $L_1=10^5$. Panels C and F show the mean Tajima's
     $D$ for the same simulations.  Error bars in C and F
     are the standard error of the mean.
     }
\end{figure*}

\subsection{RHH with large values of $s$}

We have predicated Algorithm 1 on \eqref{eq:Epi} and \eqref{eq:EpiWS}, and hence it is likely that 
it will not perform adequately in parameter regimes in which \eqref{eq:Epi} or \eqref{eq:EpiWS} is 
not accurate. Equation \eqref{eq:Epi} was derived using \eqref{eq:ptau}, which used the assumption 
that $s$ is small, so it is possible that in the large $s$ regime 
\eqref{eq:Epi} will fail to accurately predict the reduction in diversity. Here, we derive a 
theoretical form that 
describes the impact of RHH in the large $s$ regime by conditioning on the 
altered dynamics of the selected locus under very strong selection.  

For genic selection, the dynamics of the selected locus are described by

\begin{equation}
\label{eq:bigs}
\frac{d}{dt} x(t) = \frac{sx(t)(1-x(t))}{1+2sx(t)}
\end{equation}
\noindent
For small $s$, the denominator of \eqref{eq:bigs} is very close to 1 and is typically ignored (and 
was ignored in the derivation of \eqref{eq:ptau} by SWL). However, for very large $s$ the 
denominator is non-negligible, which slows the rate of growth of the selected site when it is at 
moderate to high frequency. To investigate RHH with large $s$, we solved 
\eqref{eq:pt} approximately, conditioning on \eqref{eq:bigs} for the dynamics of the selected site 
(see Appendix for derivation).  We find

\begin{equation}
\label{eq:ptaustar}
p^{*}_{\tau_f}(r_f) = e^{-4r_f} \left(1- \frac{2r_f}{s} \alpha^{\frac{-2r_f}{s}} 
             \Gamma\left[\frac{-2r_f}{s},\frac{1}{\alpha}\right]\right)
\end{equation}
\noindent
This result differs by only a factor of $e^{-4r_f}$ from \eqref{eq:ptau}, but makes very different 
predictions for large $s$ and $r_f$.  As $s$ increases, more distant sites can impact the
diversity at the neutral locus.  In fact, $s$ can be made arbitrarily large whereas $r_f$ is 
constrained to remain less than 0.5.  As a result, we expect that \eqref{eq:Epi} will underestimate 
the observed diversity for large $s$.  If we use Algorithm 1 to make $N$ arbitrarily small and $s$ 
arbitrarily large, \eqref{eq:ptaustar} predicts that patterns of diversity in the simulated 
population may be significantly different from the larger population because of this $s$ 
dependence.  We denote the reduction in diversity calculated with 
\eqref{eq:ptaustar} as 

\begin{equation}
\label{eq:Epistar}
E^{*}_{\alpha,s}\left[\sfrac{\pi}{\pi_0}\right] = \frac{\sfrac{r}{\lambda}}{\sfrac{r}{\lambda} + 
\alpha I_{\alpha,s}^{*}}
\end{equation}
\noindent
with an asterisk to differentiate it from \eqref{eq:Epi}. $I_{\alpha,s}^{*}$ is computed exactly as in section 2.1, but
replacing \eqref{eq:ptau} with \eqref{eq:ptaustar}.

We performed simulations of RHH with large values of $s$ to test 
\eqref{eq:Epistar}. We find that \eqref{eq:Epistar} accurately 
predicts the impact of RHH on diversity for large $s$, whereas 
\eqref{eq:Epi} is a poor predictor in the large $s$ regime (Figure 4).  We have performed this 
analysis primarily to explain the biased patterns of diversity produced by 
Algorithm 1, but we note that in some cases (e.g., microbes under extreme selection pressures),
it is possible that $s$ can be much larger than $0.1$.  Indeed, one experimental evolution study of 
\textit{Pseudomonas fluorescens} reported values of $s$ as large as $5$ and a mean value of $2.1$
(\citealt{Barrett:2006:Biol-Lett:17148371}).  If and when $s$ achieves such large values in 
recombining organisms it will be advantageous to use equation \eqref{eq:ptaustar} in place of 
\eqref{eq:ptau}.

\begin{figure*}[t]
  \includegraphics[height=380pt,keepaspectratio=true]{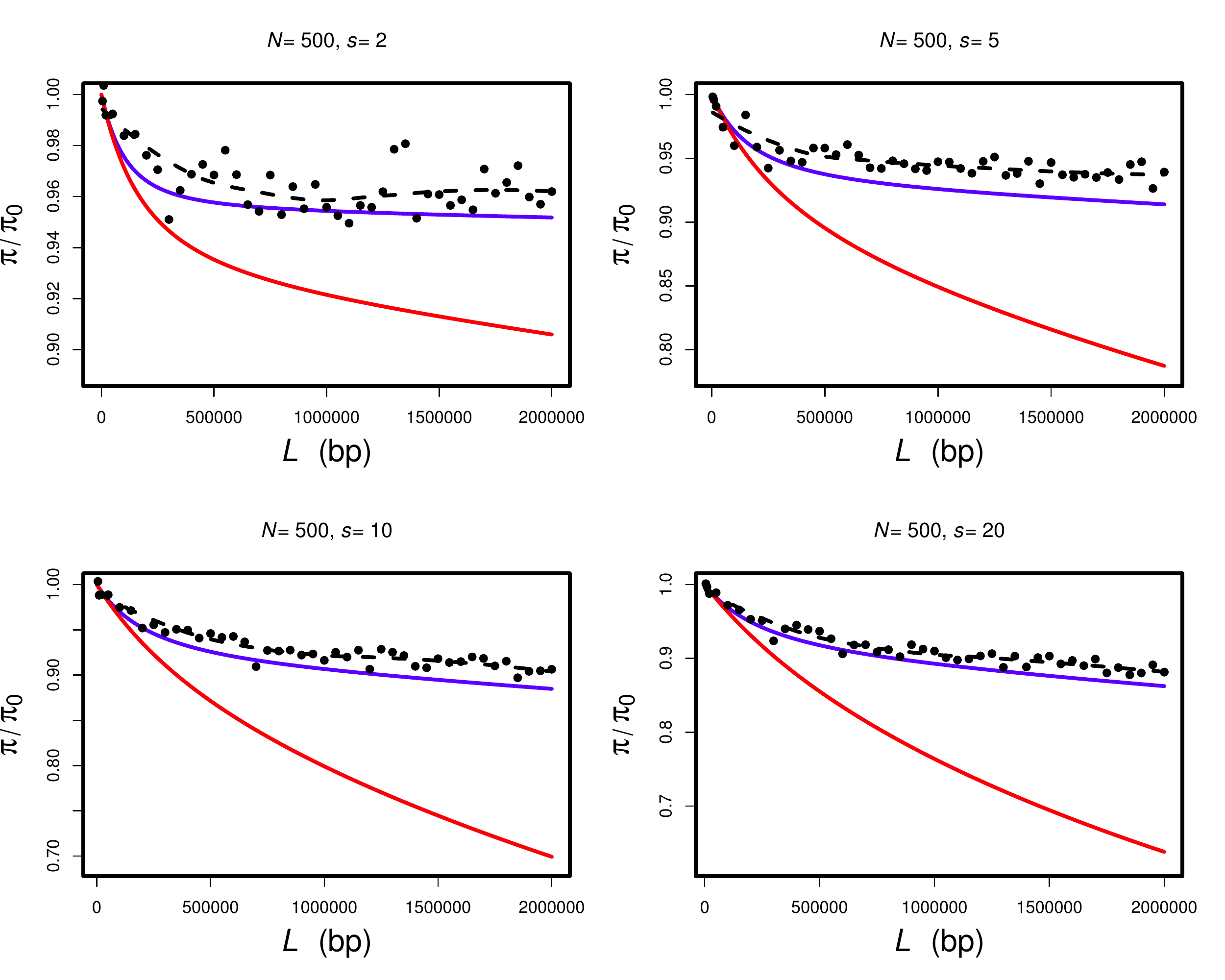}
    \caption{Simulations of recurrent hitchhiking with differing amounts of flanking sequence.  
             In each panel, we vary the amount of flanking sequence and calculate the expected 
             reduction
             in diversity using \eqref{eq:Epi} (red curves) or \eqref{eq:Epistar} (blue curves), 
             where we have used $u^{*} = \frac{2r}{s}L$ as the upper bound of the integration for
             calculating $I_{\alpha,s}$ and $I^{*}_{\alpha,s}$.  Simulation points each represent 
             the mean of 5,000 simulations, and the dashed black lines represent loess smoothing of
             the simulated data.
    }
\end{figure*}

\begin{figure*}[t]
  \centering
  \includegraphics[height=300pt,keepaspectratio=true]{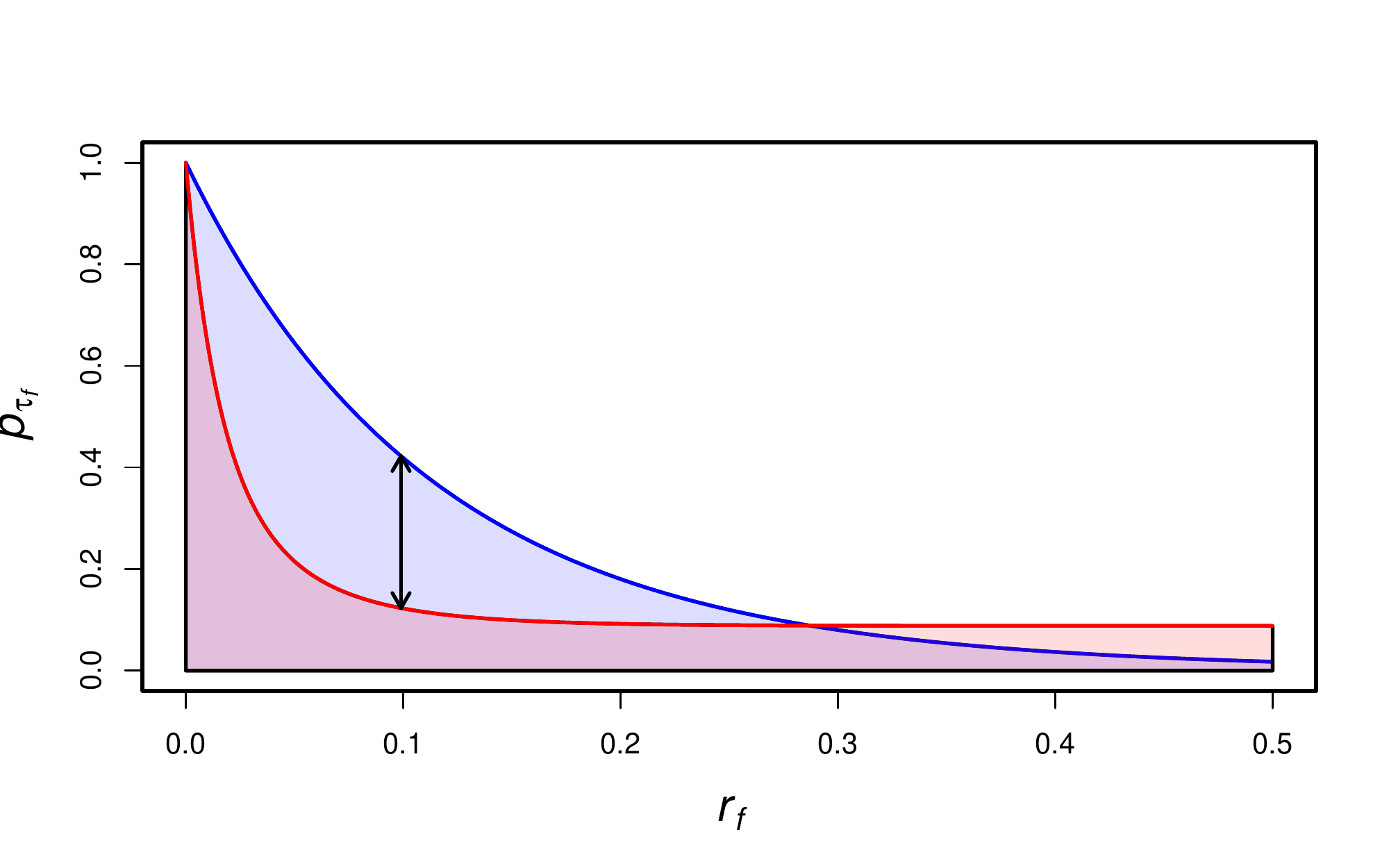}
    \caption{Algorithm 2 computes the difference between $p^{*}_{\tau_f}(r_f)$ in the population
             that we wish to model (blue) and a simulated population of smaller size (red) and
             bounds the difference in the probability of identity in the simulated and model 
             populations.  $\delta_I$ represents the maximum difference between the area under
             the red curve and the area under the blue curve that is acceptable in simulations. 
             $\delta_p$ represents the maximum difference between the red and blue curves that 
             is acceptable over all values of $r_f$ (represented by the black arrow).  
             Qualitatively, $\delta_I$ constrains the overall diversity in the simulated 
             sequences, while $\delta_p$ constrains the shape of the probability of
             common ancestry during a sweep as a function of recombination distance. 
     }
\end{figure*}

\subsection{Robust parameter rescaling for RHH simulations}

Using the results in the previous section, we modify Algorithm 1 to 
guard against violating the assumptions of the RHH model as we rescale the 
parameters of the simulations.

Let $N_1$, $L_1$, etc., be defined as in the previous section.  Our goal is to reduce $N_0$ as much 
as 
possible without altering the underlying dynamical process by more than a prespecified amount. 
Let $\delta_I$ be the maximum deviation between $I^{*}_{\alpha,s}$ for a population of size $N_1$ 
and the model population of size $N_0$ that we are willing to accept in our simulations.  For 
example, let $\delta_I=0.01$ if we desire simulated sequences in which $I^{*}_{\alpha,s}$ in a 
population of size $N_1$ differs by no more than 1\% from a population of size $N_0$. Let 
$\delta_p$ be the maximum difference between $p^{*}_{\tau_f}(u)$ in populations of size $N_0$ and
$N_1$ that we are willing to accept in our simulations, over all $u$ from $0$ to $u^{*}$, the 
length of the flanking region.  Qualitatively, $\delta_I$ is a constraint on the total area under 
$p^{*}_{\tau_f}(u)$, which influences the overall level of diversity, while $\delta_p$ is a 
constraint on the shape of $p^{*}_{\tau_f}(u)$, which influences the coalescent dynamics for a
substitution that occurs at a given distance from the neutral locus.  See Figure 5 for a pictorial
explanation.  We formalize these constraints in Algorithm 2.  Note that every parameter chosen by 
Algorithm 2 is exactly with consistent Algorithm 1, but in Algorithm 2 we precompute how small
we can make $N$ without altering the dynamics of the RHH model.

\begin{algorithm}
\begin{algorithmic}
\Procedure{Algorithm 2}{$\rho_0,\alpha_0,L_0,N_0,\lambda_0,L_1,\delta_I,\delta_p$}
\State Let $s_0 = \frac{\alpha_0}{2N_0}$; $r_0 = \frac{\rho_0}{4N_0}$; $a = \frac{s_0}{L_0r_0}$; $u_{max} = \sfrac{2}{a}$
\State $\alpha_1 = \alpha_0 = \alpha$
\State Numerically solve $\frac{I_{\alpha,s0}-I_{\alpha,s1}}{I_{\alpha,s0}} = \delta_I$ for the quantity $s_1$.
\State $D$ = Max[$p^{*}_{\tau_f,s_0}(u)-p^{*}_{\tau_f,s_1}(u)$] over all $u$ on [0,$u_{max}$]
\If{$D > \delta_p$} 
     Numerically solve Max[$p^{*}_{\tau_f,s_0}(u)-p^{*}_{\tau_f,s_1}(u)$]=$\delta_p$ for $s_1$ over all $u$ on [0,$u_{max}$] 
\EndIf
\State $N_1 = \frac{\alpha_1}{2s_1}$
\State $r_1 = \frac{s_1}{aL_1}$
\State $\lambda_1 = \frac{r_1\lambda_0}{r_0}$ 
\State $\rho_1 = 4N_1r_1$
\State \textbf{return} $N_1,\rho_1,\alpha_1,L_1,\lambda_1$
\EndProcedure
\end{algorithmic}
\end{algorithm}

We implemented Algorithm 2 in Python, using the numerical optimization tools in SciPy 
(\citealt{scipy}) for the 
numerical optimization steps. In Figures 3D-F, we demonstrate the performance of Algorithm 2 for three 
different values of $\delta_I=\delta_p$. Smaller values of $\delta$ 
generate sequences that are more closely matched to the diversity in the population of size $N_0$, 
but require a larger simulated $N_1$ and are hence more computationally intensive.  Note that
for the values of $\delta$ that we have chosen herein, only a very small change in the overall
diversity is expected.  While larger values of $\delta$ may be acceptable for some applications,
we do not recommend large values in general because the underlying dynamics are not necessarily 
expected to be conserved even if the change in overall diversity is small. Indeed, small deviations in 
mean $\pi$ and Tajima's \textit{D} are observed for the largest value of $\delta$ with strong selection (3F). 

Computational performance of the rescaled simulations is shown in Figure 11 (see Appendix). 

\subsection{The notion of ``sufficiently distant'' flanking sites}

We designed Algorithm 2 to work for any given $L_0$ in a population of size $N_0$.  In the RHH 
literature, flanking regions of $L_0=\frac{s}{r}$ are of particular interest because equation 
\eqref{eq:ptau} suggests that sites that are more than $\frac{s}{r}$ base pairs from the neutral 
locus have no impact on the neutral site \citep{Jensen:2008:PLoS-Genet:18802463}, at least for 
small $s$.  
However, this result does not hold in the large $s$ regime.  
First, the recombination fraction is not linear in the number of base pairs of flanking sequence 
when $r_f>0.1$, and $r_f$ cannot exceed $0.5$, even for arbitrarily long flanking sequences.  
Equations \eqref{eq:ptau} and \eqref{eq:ptaustar} are functions of 
$\frac{r_f}{s}$, so as $s$ gets large ($s>0.5$) it is not possible to make compensatory linear 
increases in $r_f$.  Second, the dynamics of the selected site are altered when $s$ is large, as we 
noted in the previous sections.  In particular, \eqref{eq:ptaustar} suggests that sites that have
$r_f=0.5$ (unlinked sites) have a non-negligible impact on the diversity at the neutral site when
$s$ is very large.  Our model predicts the impact of $L_u$ unlinked sites to be

\begin{equation}
\label{eq:unlink}
E[\sfrac{\pi}{\pi_0}] = \frac{1}{1+2N\lambda L_u\ p^{*}_{\tau_f}(r_f=0.5)}
\end{equation}

\begin{figure*}[t]
  \includegraphics[height=190pt,keepaspectratio=true]{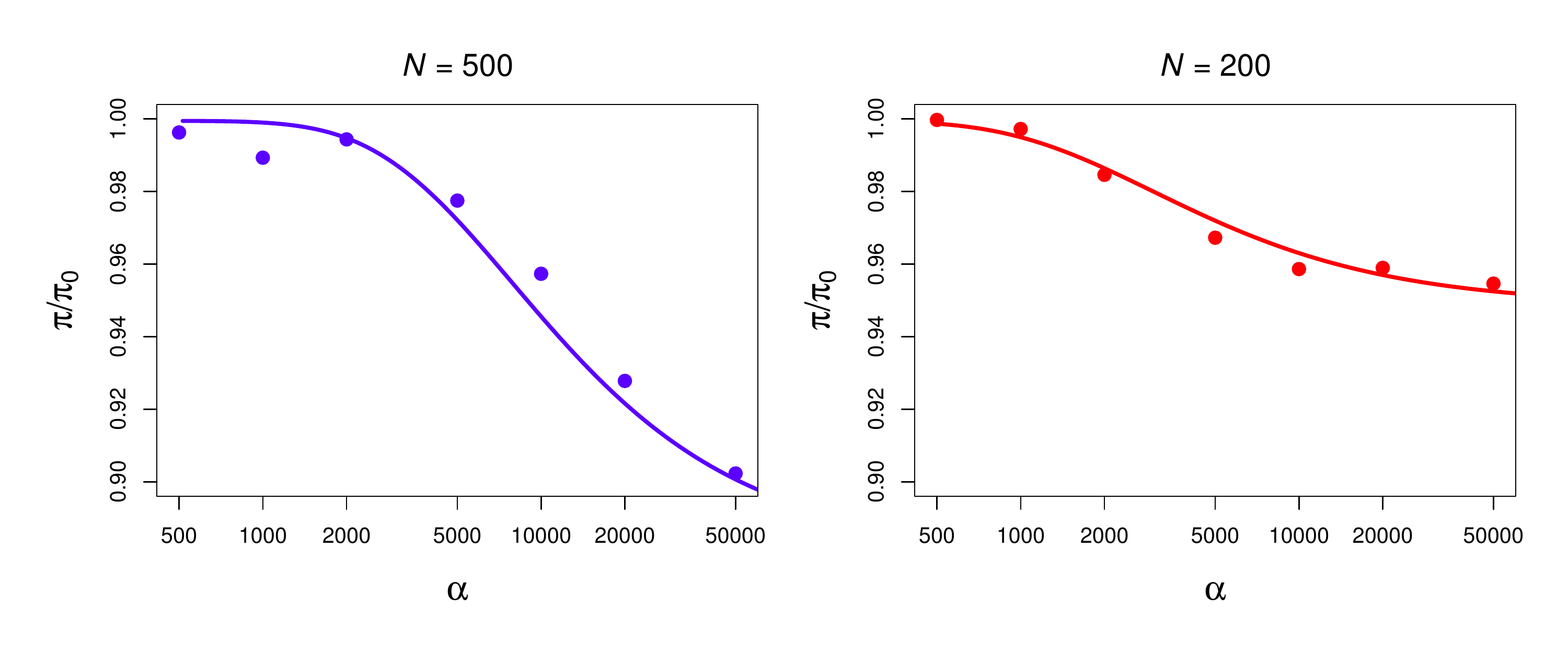}
    \caption{Simulations of RHH including only a neutral region and $L_u$ explicitly unlinked 
             selected 
             sites. $L_u=2000$, 
             $\lambda=0.5 \times 10^{-6}$. Points represent the mean of $4\times 10^4$ simulations, 
             solid lines were calculated with equation \eqref{eq:unlink} in \textit{Mathematica}. 
     }
\end{figure*}

Figure 6 shows the reduction in diversity, relative to neutrality, for simulations of RHH that 
include a neutral region and $L_u$ unlinked selected sites, and
\textit{no} linked selected sites. Equation \eqref{eq:unlink} accurately predicts the reduction in 
diversity for these simulations.  These results highlight 
another problem with Algorithm 1.  In Algorithm 1, we linearly increase the flanking sequence as 
$s$ increases. However, for large $s$, the majority of these flanking sites are essentially 
unlinked to the neutral locus, but can have a non-negligible impact on the neutral locus.  This
is fundamentally different from the dynamics in the small $s$ regime, where unlinked sites
have no impact on the neutral locus. 

While this result may not be intuitive, it is a natural consequence of very large values of $s$.  Consider 
the implausible but instructive case when $s \approx 2N$.  In the first generation after the 
selected site is introduced into the population, approximately half of the offspring are expected to be 
descendants of the individual with the selected site. At a locus that is unlinked to the selected
site, one of the two chromosomes of the individual with the selected mutant is chosen with equal probability 
for each of the descendants, which causes an abrupt and marked decrease in diversity.  Though this effect is more subtle in 
our simulations in Figure 6 (which have $1 < s \ll 2N$), there is a measurable decrease in $\pi$ due to the accumulated 
effect of unlinked sites with large $s$.

\subsection{The role of interference}

In the previous sections, we have restricted our analysis to parameter regimes in which interference between
selected sites is very rare, which is an assumption of the RHH model.  However, one of the advantages of forward 
simulations is that they can be performed under conditions with high levels of interference.

\begin{figure*}[t]
  \includegraphics[height=138pt,keepaspectratio=true]{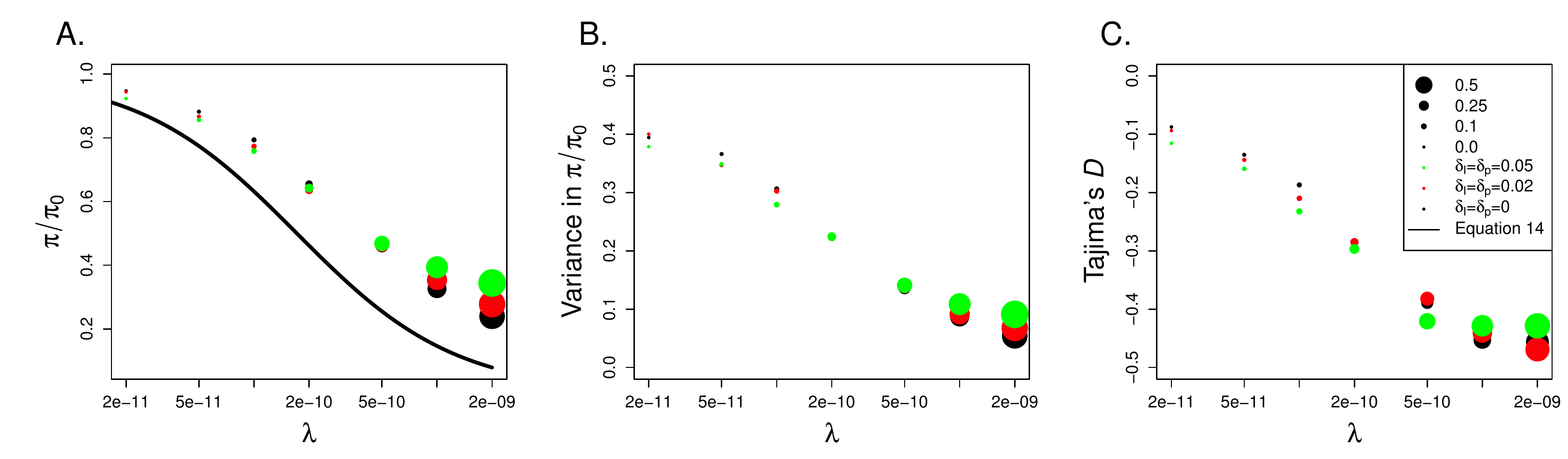}
    \caption{Rescaled simulations of RHH with increasing amounts of interference.  10,000 simulations were performed
             for each data point.  Point sizes are indicative of the amount
             of interference in the simulations, as measured by the fraction of selected substitutions that overlap
             with at least one other positively selected substituted allele while both are segregating in the population. 
             Parameters: $\alpha=2 \times 10^3$, 
             $L_0=\frac{s_0}{r_0} =4 \times 10^6$, $r_0 = 5 \times 10^{-8}$, $N_0 =5 \times 10^3$.
     }
\end{figure*}

In Figure 7, we examine the performance of Algorithm 2 with very high rates of positive selection.  Note that
the value of $\lambda$ on the $x$-axis is the expected value in the absence of interference in a population of size $N_0$, 
and the observed value of $\lambda$ in the simulations is slightly lower. 
Point sizes in Figure 7 indicate the amount of interference between selected sites, as measured by the fraction of 
selected substitutions
that overlap with at least one other substitution while segregating in the population.  This is a conservative metric for the
total effect of interference because it does not include the fraction of selected sites that are lost due to competition with
other selected sites.

As the rate of interference increases, the theoretical predictions of equation \eqref{eq:ptaustar} underestimate 
the reduction in diversity by an increasing amount (7A, black points).  This is expected because as interference increases, 
a smaller
fraction of selected sites reach fixation, and furthermore the trajectories of the sites that fix are altered due 
to competition.  Neither of these effects is modeled by equation \eqref{eq:ptaustar}.

More strikingly, as the rate of interference increases, the separation between the rescaled populations (green and red
points) and the original population (black points) also increases.  This demonstrates that Algorithm 2 does not 
recapitulate the expected diversity in the rescaled populations when the rate of interference is high in the population 
of size $N_0$.  This result is 
expected when we consider that the rate of interference is a function of both the rate of substitutions and the
time that selected substitutions segregate in the population before fixation.  It is well known that the time to fixation
is a function of both $\alpha=2Ns$ and $N$, and cannot be written naturally as a function of only one or the other.  Hence, 
when we rescale the population with fixed $\alpha$, we necessarily change the amount of interference.  We analyze 
this effect in more detail in the next section when we perform rescaling for two sets of parameters inferred in 
\textit{Drosophila}.

\subsection{An application to \textit{Drosophila} parameters}

In Figure 8, we perform rescaling with RHH parameters that are relevant to \textit{Drosophila}. 
\citet{Macpherson:2007:Genetics:18073425} found evidence supporting strong
selection ($s = 0.01$), which occurred relatively infrequently ($\lambda = 3.6\times 10^{-12}$)
in a \textit{Drosophila} population of size $N_0=1.5 \times 10^6$.
\citet{Jensen:2008:PLoS-Genet:18802463} found weaker ($s = 0.002$), more frequent selection
($\lambda = 10^{-10}$) in a population of $N_0 = 10^6$.  
Our goal is not to debate the ``true'' parameters, but rather to investigate
the practicability of rescaling using previously inferred parameters. Assuming a flanking sequence length of
$L_0=\sfrac{s_0}{r_0}$ and a recombination rate of $r_0=2.5\times10^{-8}$, we apply Algorithms 1
and 2 to these parameter sets and investigate the effect on diversity.

\begin{figure*}[t]
  \includegraphics[height=380pt,keepaspectratio=true]{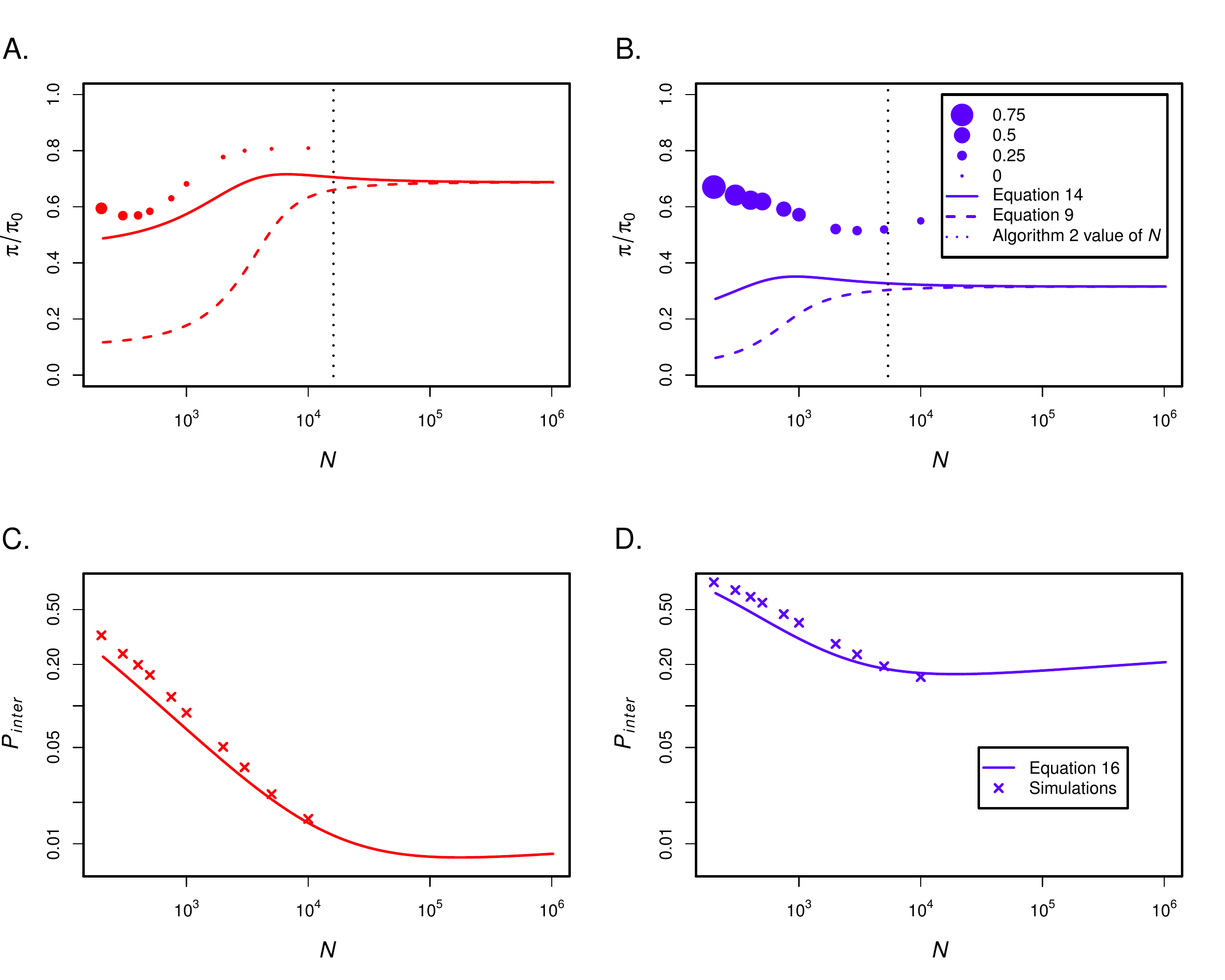}
    \caption{The reduction in diversity under previously inferred RHH parameters. Panels A and C
             use the parameters inferred by \citet{Macpherson:2007:Genetics:18073425} while
             panels B and D use the parameters inferred by \citet{Jensen:2008:PLoS-Genet:18802463}.
             The solid curves in A and B are given by \eqref{eq:Epistar},
             and the dashed curves are given by \eqref{eq:Epi}.
             Panels C and D show the probability of interference as measured by the proportion of
             substitutions that are introduced while another substitution is segregating in
             the population.  The solid curves are given by \eqref{eq:pinter}.  The size of the
             points in A and B is proportional to the observed value of $p_{inter}$, which plotted in
             C and D. The dotted vertical lines show the value of $N$ where $\delta_I=\delta_p=0.05$.
   }
\end{figure*}

In Figure 8A, we show that the trend in simulated diversity (solid red curve) as a function
of $N$ under the
\citet{Macpherson:2007:Genetics:18073425} parameters is correctly described by \eqref{eq:Epistar},
which predicts that the diversity decreases at low $N$.  However, the model slightly underestimates
the mean diversity compared to simulations.  In contrast, the model predictions of the diversity
are 
very inaccurate under the parameters estimated by \citet{Jensen:2008:PLoS-Genet:18802463} (Figure
8B, solid blue curve).  
In both 8A and 8B, Algorithm 1 strongly alters the patterns of diversity as $N$ is decreased. 
The value of $N_1$ calculated with Algorithm 2 and $\delta_I=\delta_p=0.06$ are shown with the dotted
vertical line.

Point sizes in 8A and 8B indicate the proportion of substitutions that are introduced while
another substitution is on the way to fixation (as in Figure 7).
While the interference is fairly mild for large values of $N$ under the parameters of
\citet{Macpherson:2007:Genetics:18073425}, the amount of interference is
extreme at all values of $N$ under the \citet{Jensen:2008:PLoS-Genet:18802463} parameters.  In
both cases the amount of interference in the simulations changes as we rescale $N_1$.

We designed Algorithm 2 under the assumption that interference is negligible in the population of
size $N_0$.  This assumption is approximately met under the parameters of
\citet{Macpherson:2007:Genetics:18073425}, where sweeps are infrequent and overlapping sweeps are
rare.  However, this assumption is broken by the parameters of
\citet{Jensen:2008:PLoS-Genet:18802463}, where the rate of sweeps is more than an order of
magnitude higher.  As a result, the diversity is not well predicted by equation \eqref{eq:Epistar}
at any value of $N$, and Algorithm 2 fails to generate sequences with accurate patterns of genetic 
diversity.

In general, it is useful to know \textit{a priori} when the assumptions of the RHH model are not met
as a result of high interference in a population of size $N_0$.  Consider the probability that a
positively selected substitution arises in the population while another substitution is heading
towards fixation, $p_{inter}$.  Under the assumption that interference is sufficiently infrequent
such that the mean time to fixation and probability of fixation are not strongly altered, we can
approximate $p_{inter}$ as

\begin{equation}
\label{eq:pinter}
p_{inter} \approx 1 - \left(1-2L\lambda\right)^{\tau_f} = 1 - \left(1-\frac{2s\lambda}{r}\right)^{\frac{2(1+s)\log{2N}}{s}} \approx \frac{4(1+s)\lambda\log{2N}}{r}
\end{equation}
\noindent

The probability of no substitution in a single generation
is $1-2L\lambda$, and hence the probability that no new selected substitutions are introduced while a
given selected mutation is on its way to fixation is $\left(1-2L\lambda\right)^{\tau_f}$. Supposing that $L = \frac{s}{r}$ 
and plugging in the expectation of $\tau_f$ garners the rest of the terms in the equation. 
The final approximation is valid for very small $\frac{2s\lambda}{r}$.
Note that we
do not expect \eqref{eq:pinter} to hold exactly in any parameter regime because the time to
fixation is actually a random variable (and furthermore, both $\tau_f$ and $\lambda$ are altered when interference is 
frequent), but we find that $p_{inter}$ is a useful approximation for describing the interference in simulations during 
the rescaling process.

In Figures 8C and 8D, we investigate the behavior of \eqref{eq:pinter} as we rescale the population size.  
As $N$ is decreased from $N_0$, the value of $p_{inter}$ initially decreases because
the product $(1+s)\log{2N}$ decreases.  However, as $N$ gets very small with Algorithm 1,
$(1+s)\log{2N}$ eventually begins to increase because $1+s$ increases faster than $\log{2N}$
decreases, increasing the amount of interference.

Although the exact calculation of the effects of interference is very challenging, it is
straightforward to calculate the value of $p_{inter}$ in a population of size $N_0$.  If the
value of $p_{inter}$ is large (e.g., $>0.05$, as with the parameters in Figure 8B), then the assumptions of Algorithm 2 
are broken and
there is no guarantee that the rescaled simulated sequences will be sufficiently accurate.  By contrast, if there is low 
interference
in a population of size $N_0$ then it is safe to perform rescaling so long as the value of $p_{inter}$ is constrained.
In practice, the value of $p_{inter}$ (or other quantities that are related to the rate of interference) can
be taken as an additional constraint in the calculation of $N_1$ in Algorithm 2 such
that the impact of interference is limited under rescaling.

\section{Discussion}

Simulations are an integral part of population genetics because it is often difficult to obtain 
exact analytical expressions for many quantities of interest, such as likelihoods for sequence data
under a given model.  Until recently, forward simulations were not practical because of the 
large computational burden that they can impose. However, several new forward simulation
techniques have been proposed and published 
(\citealt{Hoggart:2007:Genetics:17947444,Hernandez:2008:Bioinformatics:18842601,
Zanini:2012:Bioinformatics:23097421,
Aberer:2013:BMC-Bioinformatics:23834340,Messer:2013:Genetics:23709637}), 
and their use in population genetic studies is becoming increasingly popular.

Despite these computational advances, it remains very computationally intensive to simulate large 
populations and long chromosomes in a forward context.  It is frequently necessary to perform 
parameter rescaling to achieve computational feasibility for parameter regimes of interest 
(e.g., $N>10^5$ with long flanking sequences), particularly for applications such as 
Approximate Bayesian Computation which 
require millions of simulations for accurate inference. The hope of such rescaling efforts is that 
expected patterns of diversity will be maintained after rescaling, and that the underlying 
genealogical process will remain unaltered.

In this investigation, we tested a ``naive'' approach to
parameter rescaling, and showed that this approach can strongly alter the expected patterns of
diversity because it does not conserve the underlying genealogical process at the neutral site.
In particular, for fixed values of $\alpha$, $s$ can get arbitrarily large as $N$ is decreased, and
previous theoretical results do not accurately predict the patterns of diversity in 
this parameter regime. We derived a new theoretical form for the reduction in diversity when $s$ 
is large, and show that it has strong predictive power in simulations. We leveraged this 
result to develop a simple rescaling scheme (Algorithm 2) that approximately conserves the 
underlying genealogical process.  We note that in practice Algorithm 2 may not always be necessary, 
and as long as $s$ remains small (say, $<0.1$) Algorithm 1 will suffice.  The advantage of 
Algorithm 2 is that it allows us to quantitate the effect of rescaling, and to get the best 
possible computational performance for a given error tolerance.

It will be of great interest to extend the rescaling results for recurrent sweeps
presented herein to models that include arbitrary changes in population size and interference
between selected sites.  In the case of changes in population size, we note that the strategy 
presented in Algorithm 2 can be easily extended to perform optimization across a range of
population sizes such that the constraints are simultaneously satisfied at all time points in 
a simulation.  This strategy is consistent with previous approaches to rescaling in the 
context of complex demography \citep{Hoggart:2007:Genetics:17947444}.

Interference poses a greater challenge, because the amount of interference is
dependent both on the rate of substitution and the time that selected sites segregate. It is well
known that the time to fixation of selected alleles cannot be written as a simple function of 
$\alpha$ and depends on $N$ as well, and hence rescaling population size with fixed $\alpha$
alters the effects of interference on sequence diversity \citep{Comeron2002}. Improved understanding of scaling laws for interference 
may be necessary in order to develop appropriate rescaling strategies, to the extent that such
rescaling is possible at all in a forward simulation context. 
However, recent progress was made in the case of very strong interference, where scaling
laws were recently derived by \citet{Weissman:2012:PLoS-Genet:22685419}.

The rescaling results presented here pose an interesting dilemma for the use of forward 
simulations in population genetic studies, as previously noted by \citet{Kim01012009}.  A major
appeal of forward simulations (as compared to coalescent simulators) is the ability to incorporate 
arbitrary models (e.g., interference between selected sites, complex demographic
processes) without knowing anything about the distribution of sample paths 
\textit{a priori}. However, if simulations are only feasible when the parameters are rescaled, 
there is 
no guarantee for any given theoretical model that the rescaling will maintain expected dynamics.  
We also note that the rescaling method proposed herein was informed by in-depth knowledge 
provided by the previous work of several authors, and that in general it may not always be obvious 
which parameters must be simultaneously adjusted to maintain expected patterns of variation in 
simulations for a given complex model.

Nonetheless, forward simulation in population genetics has a bright future. Forward simulation
remains the only way to simulate arbitrarily complex models.  For many 
populations of interest (e.g., ancestral human populations), population size is sufficiently small 
such that it can often be directly simulated without rescaling. Continued 
computational advances in both hardware and software in coming
years will expand the boundaries of computational performance of forward simulation. Finally, 
active development of the theory of positive selection, interfering selected sites, background 
selection, demographic processes, and the joint action thereof will lend further insight into 
parameter rescaling and advances in the use of forward simulations in population genetic 
studies.

\section{Appendix}

\subsection{Derivation of $p^{*}_{\tau_f}$ for large $s$}

In this section, we solve for the probability of identity $p_{\tau_f}$ when the selection 
coefficient is large.  We will be concerned with the probability of identity $p$
at various frequencies throughout the sweep process.  We will subscript $p$ with $t(x)$ to 
indicate the value of $p$ at the time when the selected site reaches frequency $x$ 
(e.g., $p_{t(\sfrac{1}{2N})}$). The trajectory of the selected site for large $s$ is given 
\eqref{eq:bigs}, while the dynamics of the neutral site are given by \eqref{eq:pt}.  We transform 
\eqref{eq:pt} into allele frequency space by dividing by \eqref{eq:bigs}.  We obtain:

\begin{equation}
\label{eq:pxbig}
\frac{d}{dx}p = \frac{(1-p)(1+2sx)}{2Nsx^2(1-x)} - \frac{2r_fp(1+2sx)}{sx}
\end{equation}

This equation can be solved in \textit{Mathematica} with the initial condition 
$p_{t(\sfrac{1}{2N})} = 1$, meaning that all backgrounds carrying the selected locus are identical 
at the neutral locus when the selected site is introduced.  At the end of the sweep, $x \approx 1$. 
We take the solution with $x = 1-\sfrac{1}{2N}$ because $x=1$ results in a singularity.

\begin{equation}
\begin{split}
\label{eq:ugly}
p_{\tau_f}&  = e^{2r_f(-2+\frac{1}{2N})+\frac{2N-2}{s-2Ns}}
        \left(2-\frac{1}{N}\right)^{-\frac{1+2Nr_f+2s}{Ns}}
        \left(\frac{1}{N}\right)^{\frac{2+\sfrac{1}{s}}{2N}} \\
      & \left(e^{\frac{2r_f}{N}}\left(\frac{1}{N}\right)^{\frac{1+4Nr_f+2s}{2Ns}}\right.
         + 4^{\frac{r_f}{s}}e^{\frac{1}{s}}\left(2-\frac{1}{N}\right)^{\frac{2+\frac{1}{s}}{2N}} \\
      & \left(
        \int_{1}^{1-\frac{1}{2N}}\frac{e^{\frac{\frac{1}{C}-
        8Nr_fsC+(1+2s)\log[1-C]-(1+4Nr_f+2s)\log[C]}{2Ns}}
        \left(-1+2sC\right)}{2Ns\left(-1+C\right)C^2}dC\ \right. \\
      & \left.\left. -\int_{1}^{\frac{1}{2N}}
        \frac{e^{\frac{\frac{1}{C}-8Nr_fsC+(1+2s)\log[1-C]-(1+4Nr_f+2s)\log[C]}{2Ns}}
        \left(-1+2sC\right)}{2Ns\left(-1+C\right)C^2}dC\right) \right)
\end{split}
\end{equation}
\noindent
Equation \eqref{eq:ugly} can be numerically integrated in \textit{Mathematica}. 
However, this solution is complicated, slow to evaluate, and provides little intuition about the 
dynamics. As an alternative, we employ an approximate solution strategy.

Following \citet{GRH:3537} and others, we subdivide the trajectory of the selected allele 
into low frequency and high frequency portions. For small $x$, the term $(1 + 2sx) \approx 1$, even 
for large $s$. As a result, there is little difference between the dynamics for small $s$ and large 
$s$ sweeps at low frequency.  We rewrite \eqref{eq:pxbig} as

\begin{equation}
\label{eq:pxsmall}
\frac{d}{dx}p = \frac{(1-p)}{2Nsx^2(1-x)} - \frac{2r_fp}{sx}
\end{equation}
\noindent
which is valid for low $x$. We define the solution to \eqref{eq:pxsmall} on the interval 
$x = [\sfrac{1}{2N}, \epsilon]$ as $p_{t(\epsilon)}$.

For $x > \epsilon$, the second term on the RHS of \eqref{eq:pxbig} dominates the first term, 
because the first term is inversely proportional to the number of selected chromosomes.  To obtain 
the high frequency dynamics of the selected allele, we take $p_{t(\epsilon)}$ as the initial 
condition and solve the following differential equation on the interval $x = [\epsilon, 1]$:

\begin{equation}
\frac{d}{dx}p = - \frac{2r_fp(1+2sx)}{sx}
\end{equation}
\noindent
We find the solution:

\begin{equation}
\label{eq:ourp}
p_{\tau_f} = \left(e^{4r_f(\epsilon-1)}\right)\epsilon^{\frac{2r_f}{s}}p_{t(\epsilon)}
\end{equation}

We can perform the exact same analysis under the assumption that the dynamics are 
given by $\frac{dx(t)}{dt} = sx(t)(1-x(t))$, as was done by SWL.  This garners the solution:

\begin{equation}
\label{eq:SWLpe}
p^{SWL}_{\tau_f} = \epsilon^{\frac{2r_f}{s}}p_{t(\epsilon)}
\end{equation}
\noindent
which differs by only a factor of $e^{4r_f(\epsilon-1)}$ from \eqref{eq:ourp}.  Since 
\eqref{eq:SWLpe} was derived under assumptions identical to those used in \citet{Stephan1992237},
we conclude that sweeps with large $s$ can be modeled with the equation

\begin{equation}
\label{eq:pbigs}
p_{\tau_f} = e^{-4r_f} \left(1- \frac{2r_f}{s} \alpha^{\frac{-2r_f}{s}} 
        \Gamma\left[\frac{-2r_f}{s},\frac{1}{\alpha}\right]\right)
\end{equation}
\noindent
This equation provides very similar results to \eqref{eq:ptau} for small $s$, 
as expected, but deviates for large $s$ (Figure 9).  

\begin{figure*}[t]
  \includegraphics[height=380pt,keepaspectratio=true]{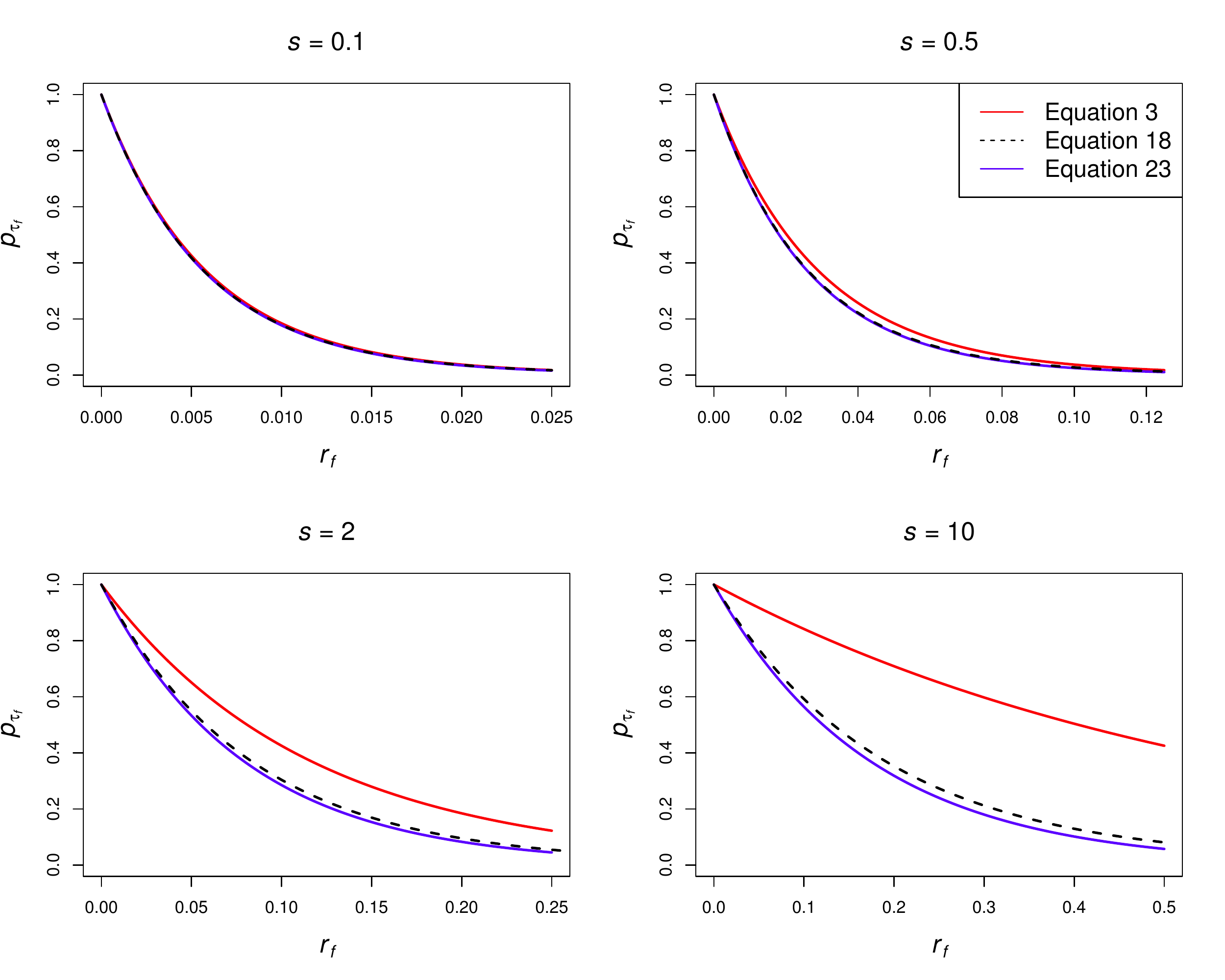}
    \caption{Equation \eqref{eq:pbigs} (blue) and Equation \eqref{eq:ugly} (black) compared to 
             \eqref{eq:ptau} (red). As expected, \eqref{eq:pbigs} and \eqref{eq:ptau}    
             are in very close agreement for small $s$, but diverge for large $s$. Equation
             \eqref{eq:pbigs} is a good approximation to \eqref{eq:ugly} across a wide range of 
             values of $s$.
             $\alpha=10^4$}
\end{figure*}

To verify that this approximation provides accurate results to the full solution given by equation 
\eqref{eq:ugly}, we compared \eqref{eq:ugly} to \eqref{eq:pbigs} in \textit{Mathematica}.  Agreement is 
very good between the exact and approximate solutions for all values of $s$ that we investigated 
(Figure 9).  

\subsection{RHH simulations in SFS\_CODE}

We performed forward simulations of RHH with SFS\_CODE
(\citealt{Hernandez:2008:Bioinformatics:18842601}). An example command line for a RHH simulation 
is:
\\ \\ \noindent
\texttt{sfs\_code 1 10 -t <$\theta$> -Z -r <$\rho$> -N <$N$> -n <$n$> -L 3 <$L$> <$l_0$> <$L$>
        -a N R -v L A 1 -v L 1 <$R_{mid}$> -W L 0 1 <$\alpha$> 1 0 -W L 2 1 <$\alpha$> 1 0 }
\\
\noindent
\\
All of these options are described in the SFS\_CODE manual, 
which is freely available online at \url{sfscode.sourceforge.net}, or by request
from the authors.  Briefly, this command line runs 10 simulations of a single population of
size $N$ and samples $n$ individuals at the end.  The recombination rate is set to
$\rho$, and 3 loci are included in the simulation.  The middle locus (locus 1) is 
$l_0$ base pairs long while the flanking loci are $L$ bp long. The middle locus is 
neutral, while the flanking loci contain selected sites with selection strength $\alpha=2Ns$. 
Every mutation in the flanking region is positively selected. The sequence is set to be non-coding 
with the option 
``\texttt{-a N r}''.  The ``\texttt{-v}'' option provides the flexibility to designate different 
rates of mutation at different loci, and the mechanics of its usage are described in detail in
the SFS\_CODE manual.  $R_{mid}$ specifies the rate at which mutations are introduced into the
middle segment relative to the flanking sequences. Please see the 
SFS\_CODE manual for a detailed example of parameter choice for RHH simulations. 

Forward simulations of DNA sequences can require large amounts of RAM and many computations.
Recurrent hitchhiking models are particularly challenging to simulate because very long sequences
must be simulated. In particular, for a given selection coefficient $s$, RHH
theory suggests that sites as distant as $r_f \approx s$ must be included in 
the simulation to include all sufficiently distant sites (see \eqref{eq:ptau}).

In many organisms, $r \approx 10^{-8}$.  Assuming $r_f \approx rL$, 
this implies that a selection coefficient of $s = 0.1$ would require $10^7$ base pairs 
of simulated sequence on each side of the neutral locus in order 
to include all possible impactful sites. This is a prohibitively large amount of sequence for
many reasonably chosen values of $\theta$ and $N$ in forward simulations (Figure 10).
However, in simulations of RHH, we are primarily interested in examining the
diversity at a short, neutral locus.  We adapted SFS\_CODE such that individual loci can
have different mutation rates and different proportions of selected sites.  For RHH simulations, we 
set the proportion of selected sites to zero in the neutral locus, and 1 in the
flanking sequence.  This greatly increases the speed and decreases RAM requirements for SFS\_CODE
because much less genetic diversity is generated in the flanking sequences (Figure 10, blue curves).
Time and RAM usage were measured with the Unix utility ``\texttt{time}'' with the command
``\texttt{/usr/bin/time -f `\%e \%M' sfs\_code [options]}''. Note that \texttt{time} reports a 
maximum resident set size that is too large by a factor four due to an error in unit conversion
on some platforms, which we have corrected herein.  Simulations were performed on the QB3 cluster 
at UCSF, which contains nodes with a variety of architectures and differing amounts of 
computational load at any given time. As such, the estimates of efficiency herein should be taken 
only as qualitative observations.

\subsection{Efficiency of rescaled simulations}

We report the time to completion of rescaled simulations relative to non-scaled populations 
using Algorithm 2 (Figure 11).  We observe reductions in time between approximately 99\% and 40\% for the parameters under consideration here.  
In general, the best performance is obtained for weaker selection, since in this case $s$ is small in the population
of size $N_0$, meaning that the value of $N$ can be changed quite dramatically without breaking the small $s$ approximation.
Better gains are also observed as the error threshold is increased, but this comes at an accuracy cost (see section 4.3).

\begin{figure*}[t]
  \includegraphics[height=240pt,keepaspectratio=true]{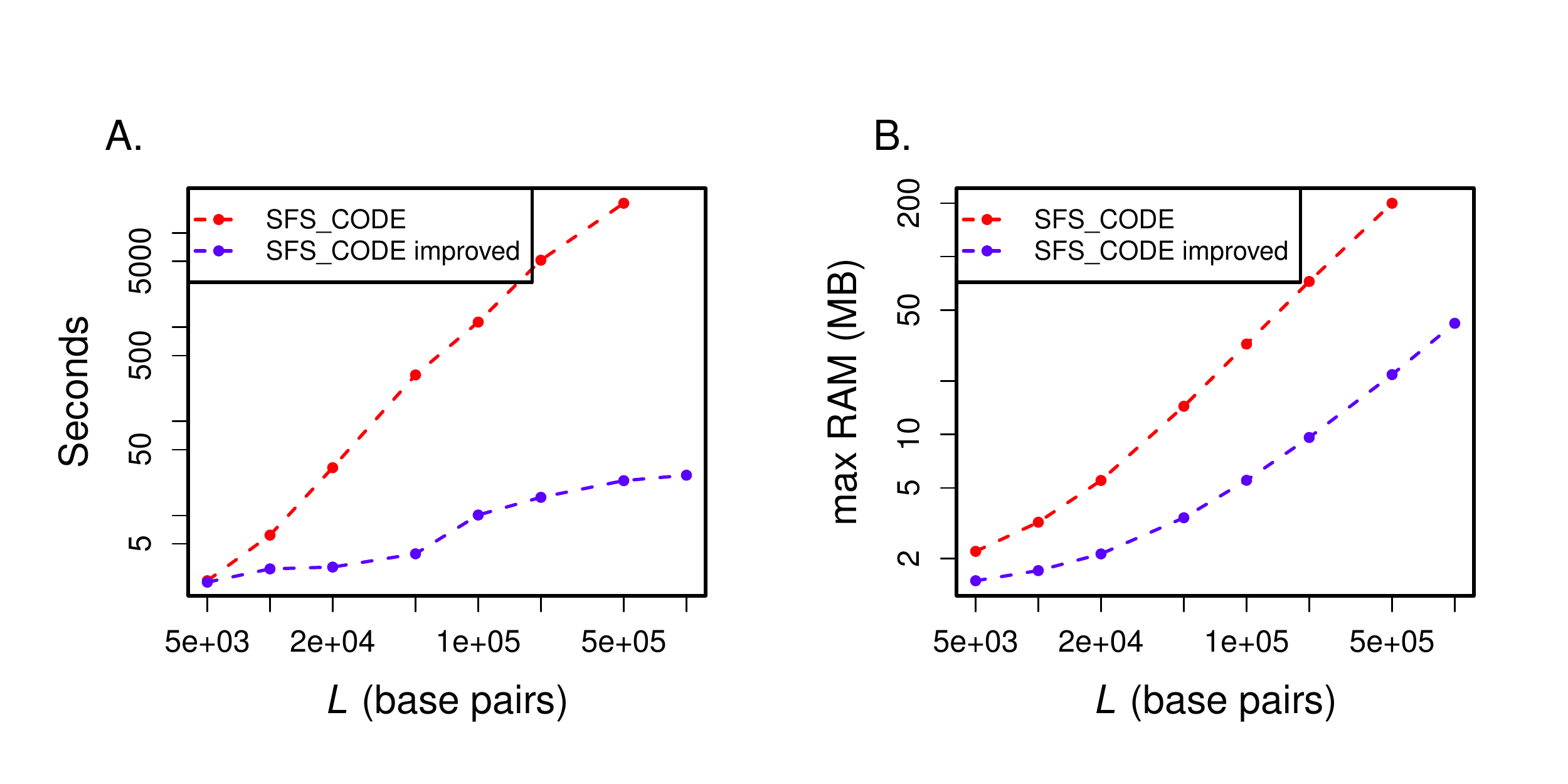}
    \caption{The computational burden of simulations of recurrent positive selection in SFS\_CODE 
             is much lower in the new version of SFS\_CODE. Both RAM requirements and time to 
             complete the simulations are reduced.  Points represent the mean of 500 simulations.
             $\theta=10^{-3}$, $\rho=10^{-3}$, $N=500$, $\lambda=10^{-9}$, $\alpha=1000$.
     }
\end{figure*}

\begin{figure*}[t]
  \includegraphics[height=240pt,keepaspectratio=true]{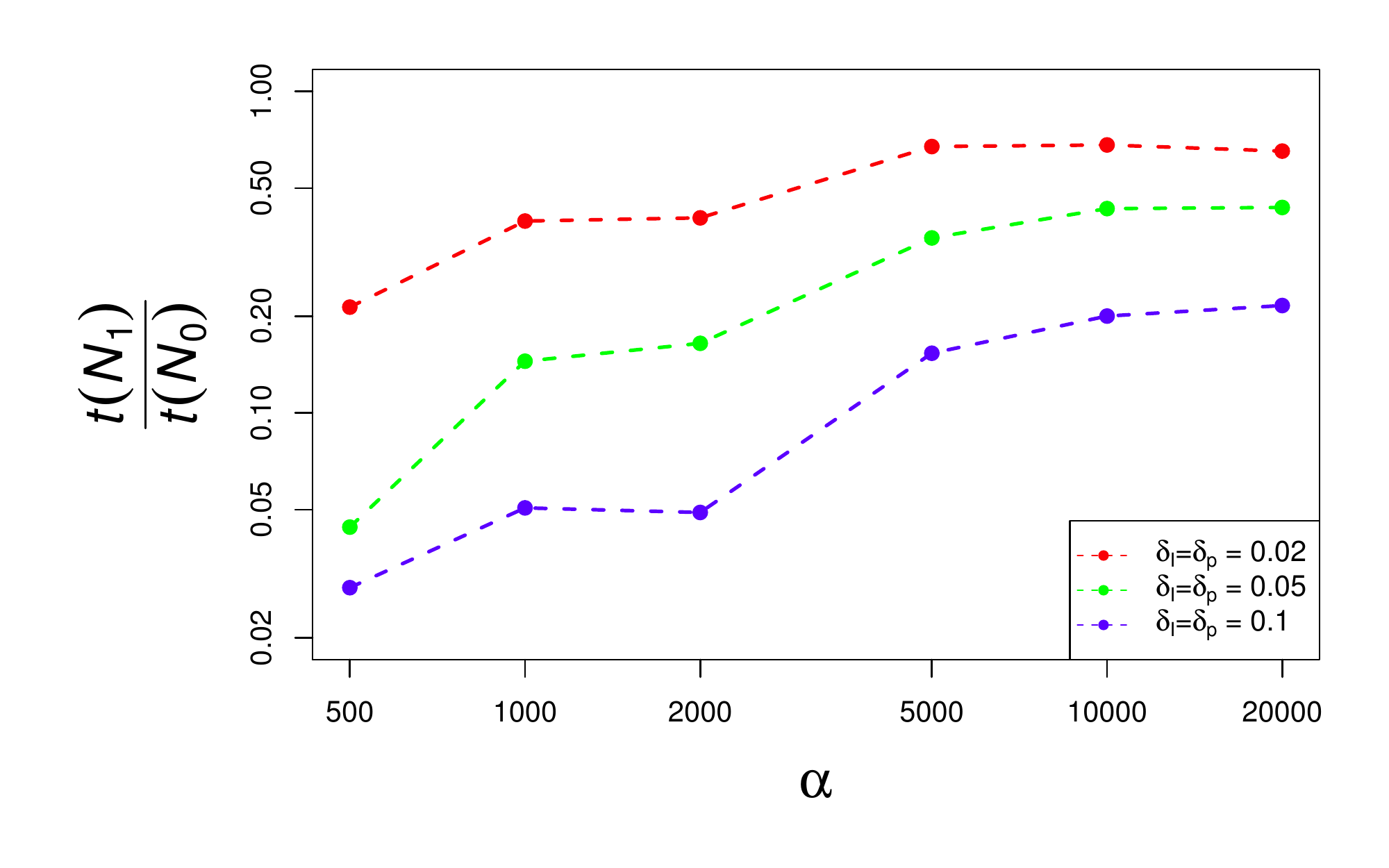}
    \caption{Duration of rescaled simulations ($t(N_1)$) relative to non-scaled simulations ($t(N_0)$).
             Parameters were chosen to match Figure 3.
             Parameters: $N_0 = 5 \times 10^3$, $\rho_0=10^{-3}$, $\lambda_0=10^{-10}$,
             $L_0=10^6$, $L_1=10^5$.
     }
\end{figure*}

\section{Acknowledgments}

This work was partially supported by the National Institutes of Health (grants P60MD006902, 
UL1RR024131, 1R21HG007233, 1R21CA178706, and 1R01HL117004-01 to R.D.H.), an ARCS foundation 
fellowship (L.H.U.), and National Institutes of Health training grant T32GM008155 (L.H.U.). We 
thank John Pool 
for stimulating discussion that motivated this research and Zachary A. Szpiech, Raul Torres, M. Cyrus
Maher, Kevin Thornton, Joachim Hermisson, and two anonymous reviewers for comments on the manuscript.

\clearpage

\bibliographystyle{genetics}
\bibliography{genetics}

\end{document}